\documentstyle[12pt,aaspp4]{article}
\lefthead{HSU \& BLAES}
\righthead{COMPTON SCATTERING}

\begin{document}

\title{Compton scattering of polarized radiation in two phase accretion disks
in active galactic nuclei}
\author{CHIA-MING HSU AND OMER BLAES}
\affil{Dept. of Physics, University of California, Santa Barbara, CA
93106}

\begin{abstract}
We present results of radiative transfer calculations of Compton
scattering of polarized radiation in a corona above an accretion disk.
The corona has a plane-parallel
geometry and is assumed to be homogeneous with a fixed temperature.
We employ a Feautrier method to calculate the radiative transfer,
incorporating the full relativistic, polarized Compton redistribution
matrix.  This allows us to explore a broad range of parameter space,
in particular large
optical depths and/or low corona temperatures, for which the
iterative scattering method is very inefficient.  Compared to Thomson
scattering, Compton scattering generally produces smaller polarizations
for featureless spectra.  Distinct frequency-dependent patterns in polarization
are produced near lines and edges, which can be used as diagnostics for
the corona.  We discuss the physics which gives rise to these results.
Compton scattering can smear out Lyman edges in total flux, and at the
same time produce large increases in polarization.  This
might help explain the steep rises in polarization observed near the Lyman
limit in some quasars.
\end{abstract}


\section{Introduction}
Active galactic nuclei (AGN) and certain X-ray binaries are commonly believed
to be powered by accretion disks.  In principle, measurement
of the polarization of radiation coming from these sources provides a possible
test for the presence of these disks.  This may be particularly
true in hard X-rays where Compton scattering is generally believed to
be the dominant radiation process.  However, until recently, there has not
been extensive discussion of Compton scattering of polarized radiation in the
literature because detection of polarization in X-rays has been very
difficult.
With the advent of the Stellar X-Ray Polarimeter on board Spectrum-X-Gamma,
as well as other X-ray polarimetry experiments, it is becoming important to
have detailed knowledge about how Compton scattering affects the polarization
of X-ray radiation in accretion disk models.

This problem has also become interesting in view of the role that Compton
scattering may play in smearing the Lyman edge in quasar accretion disks
(Czerny \& Zbyszewska 1991; Lee, Kriss, \& Davidsen 1992).
Such an effect might help explain why Lyman edges are generally
not observed, either in absorption or emission, in quasar spectra (Antonucci,
Kinney, \& Ford 1989; Koratkar, Kinney, \& Bohlin 1992).  On the other hand,
steep rises in polarization near the Lyman limit have been
observed in some quasars (Impey et al. 1995, Koratkar et al. 1995).  It may
be that these result from scattering and absorption opacity in cool disks
(Blaes \& Agol 1996, but see also Shields, Wobus \& Husfeld 1997).  If so,
then one might ask whether Compton scattering
could smear the edge out in flux but preserve a rise in polarization.

It is the purpose of this paper to address these issues and to explore some of
the basic physical effects of Compton scattering on polarization in the optical,
ultraviolet, and X-ray portions of the spectrum.  We focus
primarily on the scattering from a corona above the disk, and neglect the
reflection component above $\sim10$~keV.  This latter component is in fact
likely to be polarized itself (Matt 1993).
Similarly, we focus exclusively on the polarizing
effects of the corona, and neglect reprocessing of Compton scattered radiation
by the disk (cf. Matt, Fabian, \& Ross 1993).

This paper is organized as follows.  In section 2, we describe our radiative
transfer code which fully incorporates
relativistic effects on scattering (cf. Poutanen \& Svensson 1996, hereafter
PS96). Then we
present results for idealized continuum models in section 3, followed by
models with lines and edges in section 4.  In section 5, we present spectra and
polarization of Comptonized radiation near the Lyman limit from more realistic
accretion disk atmospheres.  Finally, in section 6 we summarize
our conclusions.  All spectra presented in this paper are
local to the disk/corona system, i.e. we neglect the relativistic transfer
function from the disk to the observer at infinity.

\section{The Numerical Method}
We use the simplest two-phase accretion disk model, in which the hot corona is
a homogeneous, plane parallel slab of electrons and ions
above a cold accretion disk (e.g. Haardt \& Maraschi 1991).
The only physical process we consider inside the corona is
Compton scattering, and, in particular, we neglect bremsstrahlung and pair
production.
Since the corona is hotter than the radiation emerging from the disk,
electrons will in general lose energy
to the radiation field.  We assume that the cooling of the corona
due to Compton scattering is immediately compensated by non-radiative
heating, so that the temperature of the corona is fixed.
As a result of the plane parallel geometry, the radiative
transfer equation has the following one dimensional form:
\begin{equation}
    \mu\,{d{\bf I}(\tau ,x,\mu)\over d\tau}=\sigma(x){\bf I}(\tau ,x,\mu)-
    {\bf S}(\tau ,x,\mu),
\end{equation}
where $d\tau = - \sigma_{T}n_{e}dz$ is the differential Thomson optical depth,
$z$ is the vertical height measured from the base of corona,
$x\equiv h\nu/m_{e}c^{2}$ is the dimensionless photon energy,
$\mu=\cos\theta$, $\theta$ is the angle between the radiation propagation
direction and the normal to the accretion disk plane, and $\sigma(x)\equiv
\sigma_{CS}(x)/\sigma_{T}$ is the ratio of the thermal Compton scattering
cross-section to the Thomson cross-section.
The disk and corona are not expected to contain significant
sources of circular polarization, and we neglect
Faraday rotation completely in this paper (see, e.g. Agol, Blaes, \&
Ionescu-Zanetti 1997).  Symmetry therefore implies that the radiation field
can be polarized in only two directions: parallel to the disk plane (for which
we choose the Stokes parameter $Q>0$), or
in a plane containing the radiation propagation direction and
the normal to the disk plane ($Q<0$).
The polarized radiation field can be fully described
by a Stokes vector ${\bf I}(\tau ,x,\mu)\equiv\left({I\atop Q}\right)$,
from which the degree of polarization is calculated as $P=Q/I$.
The source function
${\bf S}(\tau ,x,\mu)$ has the following form (Nagirner \& Poutanen 1993):
\begin{equation}
\label{eqsourcefunc}
    {\bf S}(\tau ,x,\mu)=x^{2}\int_{0}^{\infty}{dx_{1}\over x_{1}^{2}}
    \int_{-1}^{1}d\mu_{1}\int_{0}^{2\pi}d\varphi_{1}
    \int_{0}^{2\pi}d\varphi{\bf \hat{R}}(x_{1},\mu_{1},\varphi_{1}\to
    x,\mu,\varphi){\bf I}(\tau ,x_{1},\mu_{1})
\end{equation}

The radiative transfer equation can be solved by the standard Feautrier method
(see, e.g., Mihalas \& Mihalas 1984), with appropriate boundary conditions.
Unless otherwise noted, all calculations in this paper assume
no external radiation illuminating the corona
from above and a perfectly absorbing accretion disk surface at the
base of the corona.
In addition to the radiation field emerging from the top of the corona, the
radiation field scattered back to the underlying accretion disk
at the base of the corona can also be solved
in the same calculation.  This not only provides information regarding the
radiation field illuminating the accretion disk surface from the corona, but
also facilitates the future possibility of coupling our results
with other numerical programs for accretion disks
to produce self-consistent numerical models for complete
disk-corona systems.  We discuss in Appendixes our discretization of
the thermal source function and other numerical details.

We have tested our code by performing calculations which reproduce results
published by independent authors.
Figure 1 shows the degree of polarization of
low frequency radiation, initially unpolarized, after Compton scattering by an
extremely low temperature
optically thick corona, plotted against the classical Chandrasekhar result of
polarization by Thomson scattering in an optically thick electron slab
(Chandrasekhar 1960).  It is clear that in the limit of low corona
temperature and low frequency radiation, Compton scattering simulated by our
code reduces to Thomson scattering, as expected.  Also shown are
calculations
reproducing polarization profiles from Thomson scattering in an optically
thin electron slab (Phillips \& M\'esz\'aros 1986, hereafter PM86).
These assume a perfectly reflecting disk surface at the corona base
as the lower boundary
condition, which corresponds to the
transparent mid-plane condition used by
PM86.  The source radiation field has a $1/\mu$ intensity profile, which
is similar to one of the cases considered in PM86.

In addition to these
calculations in the low temperature and low frequency limit, our code also
agrees with calculations of the intensity field
from a large-particle Monte Carlo simulation for Compton scattering
by coronae up to $\tau\approx 35$ (Magdziarz 1996, private communication).

To test the range of validity of our new code, we considered cases
with extremely high Thomson depth, $\tau \ge 100$.
Our code eventually fails to
produce physically reasonable results for high energy radiation.
This is due to inaccuracy in the
calculation of the Compton redistribution matrix.  In the
radiative transfer equation, the thermal Compton scattering cross-section,
which represents the absorption part of the scattering process, is calculated
independently from the redistribution matrix in the source functions.
Therefore, errors in the values of the redistribution matrix give rise
to photon number imbalances
in the scattering process.  When the optical depth of the corona is small,
errors in the final results are very small and negligible.
However, when the optical depth of the corona increases, errors will accumulate.
Eventually, they became so large that they render the results useless.  Since
the calculation of the redistribution matrix has different precision
at different corona temperatures and photon energies, the maximum optical depth
our code can handle varies with them.  For photon energies up to
10~keV, our code provides accurate results for all corona temperatures
up to 10~MeV and for scattering
optical depths from 0.2 to more than 100.

We have also written
a code using the iterative scattering method
(PS96).  Within the appropriate
operating range of the iterative scattering code, the two codes
produce similar results.

\section{Physical Effects of Compton Scattering on Continuum Polarization}

Although there have been some calculations of polarization resulting from
Compton scattering (Haardt \& Matt 1993, PS96), there has been little
discussion of the differences between thermal Compton
scattering and Thomson scattering.  In this section, we consider some
idealized problems which isolate the major physical effects.

\subsection{Optically Thick Coronae}

The polarization emerging from an optically thick, pure scattering
atmosphere was calculated long ago by Chandrasekhar (1960) in the Thomson
limit.  The polarization is independent of frequency, because Thomson
scattering is grey.  This constancy with frequency remains true for an
optically thick, Comptonizing corona, provided the emerging photon
energies are much less than $kT_{e}$ and $m_ec^2$.\footnote{The reason
is that there is then no energy scale in the scattering, and the polarization,
like the intensity $I_\nu\propto\nu^3$, must be a power law.  However, the
only power law which is bound between -1 and 1 is a constant.}~
In addition, an infinite
scattering depth ensures that the polarization will be independent of the
input spectrum at the base of the corona.  This problem therefore permits
an examination of the differences between Thomson and Compton scattering
in the simplest possible way.

Figure 2 shows polarization profiles of emerging radiation from optically
thick coronae of various temperatures.  For corona temperatures less than
$10^{7}$ K, the results are close to the Chandrasekhar (1960) calculation
based on Thomson scattering at all frequencies.  Higher corona temperatures
cause the degree of polarization to drop for all viewing angles, but it
continues to be independent of frequency.  Although the emerging intensity
is smaller for higher
corona temperatures due to upscattering of photons out of the plotted
range, it is uniform over frequency and has an
angular distribution which is indistinguishable from the Chandrasekhar (1960)
limb darkening law.
This suggests that the high temperature polarization drop arises purely
from changes in the scattering cross-section.

Consider a Cartesian coordinate system in which a monoenergetic beam of energy
$x_1$ propagates in the positive $x$-direction.  Suppose the beam is one
hundred percent linearly polarized along the $z$-axis.  Then if this beam
interacts with thermal electrons, the differential cross-section for
the total scattered intensity at energy $x$ is given from equation
(\ref{eqsourcefunc}) by
\begin{equation}
\label{eqdsidom}
{1\over\sigma_T}{d\sigma_I\over d\Omega}={x^2\over x_1^2}(\hat{R}_{11}+
\hat{R}_{12}),
\end{equation}
and that for the Stokes parameter $Q$ at energy $x$ is
\begin{equation}
\label{eqdsqdom}
{1\over\sigma_T}{d\sigma_Q\over d\Omega}={x^2\over x_1^2}(\hat{R}_{21}+
\hat{R}_{22}).
\end{equation}

Figure 3 depicts these angular distributions of scattered radiation for
thermal Compton scattering by electrons at a temperature of 1~keV.
The beam pattern depicted is for both incident
and scattered photons with the same energy of $0.1$ keV.  Note that
outgoing radiation is 100\% polarized, as expected for Thomson scattering.
This case is in fact well within the Thomson limit, but the results
exhibit significantly more forward scattering than the classical dipole
pattern.  This is a consequence of our choice of equal energy for incident
and scattered photons.  The Thomson limit of equation (\ref{eqdsidom})
is
\begin{equation}
{1\over\sigma_T}{d\sigma_I\over d\Omega}(\mu,\phi,x)={3\over16\pi x_1}
{1-\mu^2\over[\pi\Theta(1-\cos\tilde\theta)]^{1/2}}\exp\left[-{(x-x_1)^2\over4
x_1^2(1-\cos\tilde\theta)\Theta}\right],
\end{equation}
where $\tilde\theta$ is the scattering angle [$\cos\tilde\theta=(1-\mu^2)^{1/2}
\cos\phi$], and $\Theta\equiv kT_{e}/(m_ec^2)$.
When integrated over all scattered energies $x$, this reduces to the usual
dipole formula for Thomson scattering.

Figures 4 and 5
depict the beam patterns for corona temperatures of 100~keV and 10~MeV,
respectively.  It is clear that as the
corona temperature increases, the redistribution profiles become more
symmetric around the propagation direction of
incident radiation (the forward direction), and the polarized intensity becomes
smaller compared to the total intensity.  This is due to the fact that
the isotropic electron velocity distribution of the corona randomizes
the direction and polarization of scattered radiation around the forward
direction through relativistic aberration.

This explains the decrease of polarization from the Chandrasekhar
level for high temperature optically thick coronae.  Since the
scattering cross-section of linearly polarized incident
radiation becomes more symmetric around the forward direction, then if
the incident radiation is unpolarized,
the chance of photons scattering in a given direction from two perpendicular
polarization states in the incident beam is very similar.  Furthermore,
scattered radiation from purely linearly polarized incident
radiation is no longer 100{\%} polarized, as can be seen in figure 4 and 5.
This also helps reduce the polarization in the resulting spectrum.

The above
behavior of the Compton scattering cross-section should be expected in all
phenomena related to a low energy radiation field (tested up to $0.3$ keV
in our case) which is scattered by a
high temperature electron cloud.  This is not to be confused with the fact
that the Klein-Nishina cross-section for ultra-relativistic
photons in the electron rest frame becomes unpolarized for large
scattering angles (see, e.g., Berestetskii, Lifshitz, \& Pitaevskii 1982).
Although the two effects are similar in that both of them reduce
polarization of scattered radiation, one is due to the relativistic effects of
ultra-high energy incident photons, whereas the other is due to large
random motions of thermal electrons.
The reduction of polarization
by the thermal motion of electrons is less severe when the incident photon
energy is large, since then the random motions of the electrons become
small compared to the incident photon momentum.

\subsection{Continuum Polarization from Optically Thin Coronae}

Figure 6 shows polarization profiles of emerging radiation from
coronae at fixed temperature
with various scattering optical depths.  The polarization increases
with increasing optical depth.  Since the
input spectrum at the base of the corona is unpolarized, the thicker the
corona, the larger the polarization because of increased electron scattering.
Note that these results are different from PM86 (cf. Fig. 1) due to different
boundary conditions and a different input spectrum at the base of the corona.
Here we use a perfectly absorbing accretion disk surface with a uniform,
isotropic spectrum as the input.  In PM86,
they use a transparent mid-plane (reflective boundary) condition with
$1/\mu$ intensity profile sources along the corona midplane.
Their source angular distribution and the reflected radiation
enhance the anisotropy of angular distribution of the radiation field, resulting
in a higher magnitude of polarization for optically thin coronae.

Knowing the effects of Compton scattering on low energy radiation, we now turn
our attention to higher photon energies.
The behavior of Compton scattering can be seen most clearly with
unpolarized, monochromatic incident light.  In order to see
the effects of progressively higher order scattering, we show results calculated
by the iterative scattering method.  Figure 7(a) shows the
Comptonized spectrum emerging from a $\tau=0.5$, $kT_e=100$~keV corona which
is being irradiated from below by an isotropic distribution of soft (5 eV)
unpolarized monoenergetic photons.
Two viewing angles are shown, and the lowest scattering
orders are also depicted.  As is well known, the high energy part of the
spectrum is a
power law with a thermal cutoff above the electron temperature.
Since incident light is monochromatic, all radiation
away from the source frequency is scattered radiation from the corona itself.
For a moderately optically thin corona, as shown here, effective lines of
sight, from which scattered radiation can be accumulated, are longest in the
horizontal direction.  The emerging spectrum is therefore limb brightened
(except at the source frequency, see below.)

Figure 7(b) shows the corresponding degree of polarization.  As first noted
by Haardt \& Matt (1993), the polarization is initially positive near the
monochromatic source energy, where single scattering dominates the spectrum,
but then becomes negative at higher energies where the second and higher order
scatterings dominate.  For the first scattering order, the incident radiation
is the attenuated, unscattered source from the base of the corona.  This
radiation is limb darkened because the optical depth to the base of the corona
increases for more horizontal lines of sight.  It is this limb darkening
which produces the resulting positive polarization near the source energy.
Higher scattering orders produce negatively polarized radiation from limb
brightened, scattered radiation (cf. Fig. 7(a)).  This flip in
polarization between the first and second scattering orders is a strong
diagnostic.  This could be used to confirm whether or not an observed
anisotropy break in intensity is really due to the change in scattering order,
although the anisotropy break itself is most easily detectable for near
face-on viewing angles (Haardt 1993) where the polarization is weakest.

Figure 8 shows the intensity and polarization emerging from a $\tau=0.5$,
$kT_{e}=100$~keV corona with an isotropic, unpolarized,
quasi-monoenergetic ($E_{photon}\approx 5$ eV) source at the base of
the corona.  Four different viewing angles are shown.
The polarization of high energy scattered radiation varies from $-4.5$\% to
$-16.5$\%, with an edge-on view having the largest degree of polarization.
Figure 9 shows the intensity and
polarization emerging at a viewing angle of $\mu=0.069$ (edge-on view)
from $\tau=0.5$
coronae with the same source at the corona base.  This time the corona
temperature is different.  For a factor of ten decrease of corona temperature
from $kT_{e}=100$~keV to $kT_{e}=10$~keV, the
magnitude of the degree of
polarization of high energy scattered radiation
increases dramatically from $16.5$\% to $45$\%.
Figure 10 shows the intensity and
polarization emerging at a viewing angle of $\mu=0.069$ (edge-on view)
from $kT_{e}=100$~keV
coronae with the same source at the corona base.  Three different scattering
optical depths are shown, and the degree of polarization of high energy
scattered radiation varies from $-10$\% of $\tau=1.0$ to $-20$\% of $\tau=0.3$.
For the three parameters, edge-on views combined with small corona temperatures
and small scattering optical depths produce the largest degree of polarization
in a plane normal to the accretion disk surface (negative).  This can be
understood from our previous discussion.  Since thin coronae have very
anisotropic angular distributions of scattered radiation, they produce larger
degrees of polarization for edge-on views compared with thicker coronae.  And
the small corona temperature reduces the de-polarization effect of
the thermal motion of the electrons.

\section{Comptonization of Lines and Edges}

\subsection{Polarization Signature of Lines by Compton Scattering}

Figure 11 shows the intensity and polarization emerging from
a $\tau=0.5$, $kT_{e}=100$~keV corona with an isotropic,
unpolarized, continuum background and an emission line feature illuminating
the base of the corona.  The overall intensity is smaller for an edge-on
view because the edge-on view has a longer line of sight, thus more photons are
upscattered out of the plotted range.  Note that for a moderately thin
corona of $\tau=0.5$ with the continuum input shown here, the main component in
the emerging spectrum is the unscattered source.
However, scattering broadens the emission line and shifts it toward higher
energies.  For the low energy continuum below the emission line
in the original input spectrum, the emerging
polarization exhibits the behavior discussed
in Section 3.1.  The degree of polarization is independent of frequency,
and has the same value as if the input spectrum were completely uniform.
At the energy range of the emission line, the degree of polarization rises.
At energies higher than the emission line, the degree of polarization drops
below the level before the onset of the emission line, then rises again to
approach the level before the onset of the emission line from below.

The transfer equation is linear in the radiation field,
and all this behavior can therefore be explained by Compton
scattering on the continuum background and emission line
separately.  For a frequency independent continuum source, Compton
scattering produces
degrees of polarization which are independent of frequency well below
photon energies $\sim{\rm min}[kT_{e},m_ec^2]$.  For an emission line,
like the quasi-monoenergetic source in Section 3.2, Compton scattering
produces positive polarization at energies around the line and
negative polarization away from the line.  When we put both together, we
have the result in figure 11. Away from the emission line, especially
the low energy part,
the degree of polarization is largely determined by Compton scattering
effects on the uniform continuum background.  Near the emission line,
the degree of polarization exhibits
a rise when the emission line alone would
have a positive polarization and a drop when the emission line
alone would have a negative polarization.

If there is an absorption line instead of an emission line in an otherwise
uniform continuum background, the results are the reverse of the
emission line spectrum (figure 12).  As before, the overall intensity is smaller
for edge-on views.  Again, Compton scattering broadens the absorption line and
shifts it toward higher energies.
Instead of a rise then a drop in the degree of polarization toward the high
energy end of the spectrum, the absorption line causes a larger drop first at
the energy of the absorption line, then a rise at higher energies.
As we discussed in section 3.2, the first scattering order from a monoenergetic
source has positive polarization, while higher scattering orders have
negative polarization.
The emerging polarization for a uniform continuum is actually
a balance between those two opposite polarizations when all scattering orders,
including the attenuated, unscattered source, are mixed together in the final
emerging radiation.  Inside the absorption line, there is less
unscattered radiation in the original input spectrum at the base of the
corona.  This makes the corresponding unscattered and first order scattered
radiation, which have zero and positive polarization, respectively,
less important when mixed with higher order scattered radiation from outside
the line.  The result is a drop in the emerging degree of
polarization inside the absorption line due to the larger
negative polarization.  Similarly, at energies higher than the absorption
line, there is a lack of higher order scattered radiation originating
from inside the absorption line.  Therefore, a rise in the degree of
polarization is seen there due to the smaller effect of negative polarization.

The rise and fall in the degree of polarization of lines with respect to the
continuum level after Compton scattering is a generic phenomenon in
plane parallel geometry.  The magnitude
of the rise and fall in polarization for a line is affected by the
strength of the line.  Figure 13 shows the intensity and polarization
emerging along $\mu=0.069$ (edge-on view) from a $\tau=0.5$,
$kT_{e}=100$~keV
corona with an isotropic, unpolarized, continuum background plus
emission lines of varying strength
as the input spectrum at the base of the corona.  Not surprisingly,
stronger lines
have larger rises and drops in polarization, while the
polarization of the continuum stays the same, indepedent of line strength.
Note that the high corona temperature in this case keeps the polarization
small in all cases.

\subsection{Polarization of Edges by Compton Scattering}
The effects of Compton scattering on emission and absorption edges are
quite different.  In the context of the two phase accretion disk, where the
corona is hot, Compton scattering of disk radiation by electrons in the corona
tends to increase photon energy.  Hence
absorption edges have more photons scattered into the low intensity region,
which has higher energy, than emission edges.  The polarization signatures
are different, too.  Emission edges show more complex structure than
absorption edges, again due to the asymmetry of photon energy drift.

Figure 14 shows the intensity and polarization emerging from
a $\tau=0.5$, $kT_{e}=100$~keV corona with an isotropic, unpolarized
absorption edge spectrum illuminating at the base of the corona.  The intensity
is assumed to be uniform both above and below the edge.
Photons scattered into the high energy, originally low intensity region
form a power law spectrum, as expected from Compton scattering in general.
Since the scattered radiation dominates this part of the emerging spectrum,
the polarization exhibits large negative values here, akin to the high energy
spectrum of a Comptonized, monochromatic input spectrum discussed in Section
3.2.  The upturn in polarization for photon energies larger than 1~keV is due
to the fact that high order scattered radiation from frequencies lower than the
discontinuity is becoming weaker compared to radiation from
frequencies higher than the discontinuity.
For frequencies lower than the discontinuity, the polarization is
independent of frequency and has the same value it would have had if the
input spectrum were completely uniform.

Figure 15 shows the same thing for an emission edge.  There is still a power
law spectrum produced on the low energy side near the discontinuity, but
the slope is much steeper than the one in the absorption edge case.
This is because the power law in the emission edge case
is formed by downscattered photons from the high energy side of
the discontinuity, and downscattered radiation has an intrinsically steeper
slope than upscattered radiation.  The
more complex structure of the polarization in the emission edge case
can be understood by considering the two sides of the
discontinuity separately.  On the low energy side,
the emerging radiation is dominated by downscattered
radiation from the high energy side, as can be seen from the intensity plot.
Hence the resulting polarization is similar to that of the low energy
downscattered region of a monochromatic input source discussed in Section 3.2
(cf. Figs. 7-10).  The polarization is positive right next to the
discontinuity, and turns negative at lower energies.  At even lower
energy, downscattered radiation from the high energy side becomes less
important, and both the intensity and degree of polarization return to their
frequency-independent values for a uniform spectrum as discussed above.
On the high energy side of the discontinuity, the
polarization rises and then drops back toward the level for a uniform
input spectrum.  This is similar to the behavior on the high energy side of
the Comptonized absorption line discussed above (cf. Fig. 12),
and arises for exactly the same reasons.

Just as in the case of lines, the variation in polarization across the edge
depends on the edge height.  Figure 16 shows the intensity and polarization
emerging along $\mu=0.069$ (edge-on view) from a $\tau=0.5$, $kT_{e}=100$~keV
corona with an isotropic, unpolarized absorption edge spectrum
with various edge heights illuminating the base of the corona.
As expected, larger edge heights cause larger dips into negative polarization,
while the polarization away from the edge is independent of the edge height.

\section{The Lyman Edge}
Many accretion disk models of AGN predict a strong Lyman edge feature in the
spectrum, while only a very small number of observed AGN spectra possess Lyman
edge features, and even these are weak at best.
Czerny and Zbyszewska (1991) pointed out that Compton scattering
of disk radiation by a corona can smear out the Lyman edge.  We confirm
this suggestion by calculating spectra after Compton
scattering from an accretion disk atmosphere, which has a strong Lyman edge
feature (Agol 1997, private communication).  For simplicity, we assume the
corona to have uniform optical depth and temperature over the disk.

Figure 17 and 18 show the intensity and polarization entering to and emerging
from a $\tau=0.5$, $kT_{e}=100$~keV corona, with viewing angle of $\mu=0.211$
and $\mu=0.789$, respectively.  From the intensity plots, it is clear that
the sizes of the Lyman and Balmer edges are both reduced by Compton scattering,
and photons are shifted toward higher energies.
The Lyman and Balmer edges are almost smeared out by Compton scattering in
the more edge-on view in figure 17.  However, intensity discontinuities are
still evident in the more face-on view of figure 18, although they have been
reduced.

The changes in polarization caused by Compton
scattering are more subtle.  An edge-on view changes the
degree of polarization more so that
most of the polarization features disappear.  For example, in figure 17,
the polarization discontinuities at the Lyman and Balmer edges before Compton
scattering are almost smeared out, and the steep
rise in polarization blueward of the Lyman edge has completely disappeared.
Figure 18 shows that a more face-on view better preserves
both polarization discontinuities and the rise of polarization blueward
of the Lyman edge, even though the edges in total intensity are reduced.
This can be attributed to the fact that for
a corona with $\tau=0.5$, a more edge-on view has a longer effective line
of sight, and so the effect of Compton scattering is greater.
Calculations with different corona scattering optical depths show that
larger scattering optical depths produce
results similar to the more edge-on view
shown here.  Smaller optical depths have results similar to the more
face-on view, retaining more of the original features in the disk atmosphere
spectrum, both in intensity and degree of polarization.  If steep rises in
polarization blueward of the Lyman edge in several quasars as observed via HST
are actually produced in the accretion disk atmospheres (Blaes \& Agol 1996),
then this calculation implies that we should be looking at optically thin
corona with a more face-on view, if the corona exists at all.

The detailed behavior of the polarization can be understood from our
results in Sections 3 and 4.
In figure 17, the polarization for radiation frequencies lower than
the Lyman edge increases as a result of Compton scattering.
For radiation frequencies far above the Lyman edge, the original disk atmosphere
spectrum has extremely small intensity.  Therefore, higher order Compton
scattered radiation, which has a large negative degree of polarization,
dominates this part of emerging spectrum.

A major criticism against Compton scattering as a means of getting rid
of Lyman edges in AGN spectra is that the radiation should then have
substantial polarization (Antonucci 1992), at least if Compton
scattering imparted a polarization comparable to that of Thomson scattering.
We have shown in section 3 that in general Compton scattering produces
smaller polarizations than Thomson scattering for a featureless continuum.
Nevertheless, our
calculations here show that at the corona temperatures expected in AGN
($<100$~keV), if Compton scattering succeeds in smearing an edge, it
does indeed generally produce a subtantial polarization in the
optical/UV portion of the spectrum.  Hence if Compton scattering is
present, some means must be invoked to reduce the polarization
to observed values, such as Faraday rotation (e.g. Agol \& Blaes 1996).

Shields, Wobus, and Husfeld (1997) recently demonstrated that taking into
account the relativistic transfer from the black hole to a distant
observer, accretion disk atmosphere models have difficulty explaining
the steep rise in polarization blueward of the Lyman edge observed in
PG 1222+228 and PG 1630+377.  They further showed that a good fit to the
observed spectropolarimetry data can be achieved by assuming an ad hoc disk
spectrum with a sharp Lyman edge in absorption in total flux together with
a sharp jump in polarization towards high frequencies
at the Lyman edge.  Compton scattering can produce just such a spectrum.
In fact, it can even produce a steep rise in polarization {\it without}
a jump in total intensity.
In Section 4.2,
we showed that an absorption edge spectrum has large (negative)
polarization blueward of the edge after Compton scattering,
if the input spectrum at the base of the corona is unpolarized (figure 14).
With appropriate corona parameters, Compton scattering can easily produce
polarizations as high as $45$\% in a plane containing the
radiation propagation direction and the normal to the disk plane.
This is a promising
alternative for explaining the steep rise of polarization blueward of Lyman
edge observed in AGNs.

\section{Conclusions}
We have presented calculations of Compton scattering of polarized
radiation in accretion disk coronae using a finite difference Feautrier method.
Our calculations can be improved by incorporating
other sources of radiation, such as bremsstrahlung, or absorption terms.
There are also existing techniques to use the same one dimensional
radiative transfer equation on calculations of more complex corona geometry
(PS96), making our code useful for cases other than plane parallel coronae.
By calculating the radiation reflected by the corona back to the disk
atmosphere, our code can be coupled to a code that calculates the emerging
spectrum from the accretion disk atmosphere.  Iteration
between the two codes can then produce a fully self-consistent spectrum.

By considering a number of idealized input spectra at the base of the corona,
we have identified some generic features of the effect of inverse Compton
scattering on polarization.  Starting with unpolarized, monoenergetic
radiation at the base of the corona, the emerging polarization is positive
near the frequency of the unscattered source and negative further away.
When the input spectrum has uniform intensity over a range of
frequencies, radiation from different source frequencies with different orders
of scattering is mixed together to produce the emerging spectrum, with a
frequency independent degree of polarization which is determined by the
viewing angle, corona temperature, and corona scattering depth.
If there are intensity variations in the input spectrum such
as lines or edges, there will be corresponding generic frequency-dependent
patterns in the polarization, which depend on the strength and profile of the
lines or edges.

As first suggested by Czerny and Zbyszewska (1991), Compton scattering 
can be successful in smearing out the Lyman edge, provided the optical depth
along the line of sight is sufficiently high.  Generally this also smears
out any polarization features arising from the underlying disk atmosphere.
However, there are regions of parameter space where such features are
preserved even though the Lyman edge intensity jump is markedly reduced
(cf. Fig. 18).
When Compton scattering is successful in smearing out an edge, it generally
imparts a significant polarization to the overall spectrum.  This may have to
be removed by Faraday rotation, for example, in order not to conflict with
the generally low optical polarizations observed in type 1 AGN.  Finally,
Compton scattering can produce very large, negative polarizations in regions
of the spectrum where the disk itself produces very little radiation.  In this
way a steep rise in polarization can be produced near the Lyman limit, even
though the resulting Comptonized intensity is smeared out and exhibits very
little jump.

\acknowledgements

We thank E. Agol, I. Hubeny, P. Magdziarz, and A. Zdziarski for useful
conversations.  This work was supported by NSF grant AST~95-29230.

\appendix
\section{Source Functions in Finite Difference Form}

The original radiative transfer equation, as described in Section 2, is
\begin{equation}
    \mu\,{d{\bf I}(\tau ,x,\mu)\over d\tau}=\sigma(x){\bf I}(\tau ,x,\mu)-
    {\bf S}(\tau ,x,\mu),
\end{equation}
where
\begin{equation}
    {\bf S}(\tau ,x,\mu)=x^{2}\int_{0}^{\infty}{dx_{1}\over x_{1}^{2}}
    \int_{-1}^{1}d\mu_{1}{\bf \tilde{R}}(x,\mu;x_{1},\mu_{1})
    {\bf I}(\tau ,x_{1},\mu_{1})
\end{equation}
and
\begin{equation}
    {\bf \tilde{R}}(x,\mu;x_{1},\mu_{1})=
    \int_{0}^{2\pi}d\varphi_{1}\int_{0}^{2\pi}d\varphi
    {\bf \hat{R}}(x_{1},\mu_{1},\varphi_{1}\to x,\mu,\varphi).
\end{equation}
We define ${\bf j}$ and ${\bf h}$ in the usual way:
\begin{equation}
    {\bf j}(\tau ,x,\mu)={1\over 2}[{\bf I}(\tau ,x,\mu)+{\bf I}(\tau ,x,-\mu)],
\end{equation}
and
\begin{equation}
    {\bf h}(\tau ,x,\mu)={1\over 2}[{\bf I}(\tau ,x,\mu)-{\bf I}(\tau ,x,-\mu)],
\end{equation}
where $\mu$ is restricted to $1 \ge \mu \ge 0$ hereafter.

Then, the radiative transfer equation can be written as
\begin{equation}
    \mu\,{d{\bf h}(\tau ,x,\mu)\over d\tau}={\bf j}(\tau ,x,\mu)-
    {\bf S}^{+}(\tau ,x,\mu)
\end{equation}
and
\begin{equation}
    \mu\,{d{\bf j}(\tau ,x,\mu)\over d\tau}={\bf h}(\tau ,x,\mu)-
    {\bf S}^{-}(\tau ,x,\mu),
\end{equation}
where
\begin{equation}
    {\bf S}^{\pm}(\tau ,x,\mu)={1\over 2}[{\bf S}(\tau ,x,\mu)\pm
    {\bf S}(\tau ,x,-\mu)].
\end{equation}
Note that compared to many other forms of scattering, Compton scattering
is unusual in that the source function is not invariant under the transformation
$\mu\rightarrow -\mu$.  Hence there is a source function term on the right
hand side of equation (A7).
The source functions, written in terms of Compton redistribution matrixes,
are
\begin{equation}
    {\bf S}^{+}(\tau ,x,\mu)=x^{2}\int_{0}^{\infty}{dx_{1}\over x_{1}^{2}}
    \int_{0}^{1}d\mu_{1}[{\bf \tilde{R}}(x,\mu;x_{1},\mu_{1})+
    {\bf \tilde{R}}(x,\mu;x_{1},-\mu_{1})]{\bf j}(\tau ,x_{1},\mu_{1})
\end{equation}
\begin{equation}
    {\bf S}^{-}(\tau ,x,\mu)=x^{2}\int_{0}^{\infty}{dx_{1}\over x_{1}^{2}}
    \int_{0}^{1}d\mu_{1}[{\bf \tilde{R}}(x,\mu;x_{1},\mu_{1})-
    {\bf \tilde{R}}(x,\mu;x_{1},-\mu_{1})]{\bf h}(\tau ,x_{1},\mu_{1})
\end{equation}
In deriving this, we have used one of the symmetry properties of
the redistribution matrix:
\begin{equation}
    {\bf \tilde{R}}(x,\mu;x_{1},\mu_{1})={\bf \tilde{R}}(x,-\mu;x_{1},-\mu_{1})
\end{equation}
which comes from the fact that scattering processes do not depend on
the choice of coordinate orientation.

Now we have to discretize the above integrals for the source functions.
We assume that the intensities ${\bf j}(\tau)$ and ${\bf h}(\tau)$ at a
certain depth do not change dramatically within an
angle-frequency cell for a chosen finite difference grid, so that we can
approximate them as constant within each cell.
Further improvement, if necessary,
can be
introduced by using cubic spline functions to represent the intensity
${\bf j}(\tau)$ and ${\bf h}(\tau)$ (Adams, Hummer, \& Rybicki 1971).

Based on our assumption, we can then write the source functions as
\begin{equation}
	{\bf S}_{k}^{+}(\tau)= \sum_{k'=1}^{K}
    {\bf R}_{k,k'}^{j}{\bf j}_{k'}(\tau)
\end{equation}
and
\begin{equation}
	{\bf S}_{k}^{-}(\tau)= \sum_{k'=1}^{K}
    {\bf R}_{k,k'}^{h}{\bf h}_{k'}(\tau),
\end{equation}
where
\begin{equation}
    {\bf R}_{k,k'}^{j}=x_{k}^{2}\int_{x_{1,k'l}}^{x_{1,k'h}}
    {dx_{1}\over x_{1}^{2}}w_{\mu}(\mu_{1k'})
    [{\bf \tilde{R}}(x_{k},\mu_{k};x_{1},\mu_{1k'})+
    {\bf \tilde{R}}(x_{k},\mu_{k};x_{1},-\mu_{1k'})],
\end{equation}
\begin{equation}
    {\bf R}_{k,k'}^{h}=x_{k}^{2}\int_{x_{1,k'l}}^{x_{1,k'h}}
    {dx_{1}\over x_{1}^{2}}w_{\mu}(\mu_{1k'})
    [{\bf \tilde{R}}(x_{k},\mu_{k};x_{1},\mu_{1k'})-
    {\bf \tilde{R}}(x_{k},\mu_{k};x_{1},-\mu_{1k'})],
\end{equation}
$w_{\mu}(\mu_{1k'})$ is the
integration weight associated with the angular abscissas $\mu_{1k'}$ of the
$k'$ angle-frequency cell, and
$[x_{1,k'l},x_{1,k'h}]$ is the frequency range of the $k'$
angle-frequency cell.
Note that we have arranged all angle-frequency cells
into a long vector using only one index.

The radiative transfer equations then become
\begin{equation}
    \mu\,{d{\bf h}_{k}(\tau)\over d\tau}={\bf j}_{k}(\tau)-
    \sum_{k'=1}^{K}{\bf R}_{k,k'}^{j}{\bf j}_{k'}(\tau)
\end{equation}
and
\begin{equation}
    \mu\,{d{\bf j}_{k}(\tau)\over d\tau}={\bf h}_{k}(\tau)-
    \sum_{k'=1}^{K}{\bf R}_{k,k'}^{h}{\bf h}_{k'}(\tau),
\end{equation}
and we can proceed to a discretized optical depth mesh as usual in the Feautrier
method.

\section{Frequency Weights in Redistribution Matrix for Low Corona Temperature}
When the corona temperature is low, the redistribution matrix for Compton
scattering is sharply peaked in frequency since the energy shift of
photons per scattering is small.  For sufficiently low corona temperatures,
the frequency finite difference grid cannot resolve the large but narrow
variation in the redistribution matrix.  In order to correctly calculate
the frequency
weights when the redistribution matrix is narrowly peaked, without using
a large number of frequency points which would be wasteful when the
redistribution matrix is smooth, we need a versatile integration scheme.

We have devised such a scheme which works as follows.
For any frequency range
$[x_{A},x_{B}]$, we evaluate the redistribution
matrix $\bf \tilde{R}$ at the midpoint $x_{C}$ and two end points
$x_{A},x_{B}$.  Then we take
the ratio of the 11 components of the matrix between the midpoint
and whichever end point which has the larger value.  For example,
if ${\bf \tilde{R}}_{11}(x_{A}) < {\bf \tilde{R}}_{11}(x_{B})$, then the ratio
$r={\bf \tilde{R}}_{11}(x_{C})/{\bf \tilde{R}}_{11}(x_{B})$.  If the ratio $r$
is smaller than a chosen critical value (we use $10^{-5}$ or
$10^{-7}$), we flag it as a large variation in the redistribution matrix.
The routine will then perform the integration over the smaller value range
($[x_{A},x_{C}]$ in the above
example) using the trapezoidal rule, and recursively call itself to integrate
over the larger value range ($[x_{C},x_{B}]$ in the above example).
Otherwise, if the
ratio $r$ is larger than the critical value, the routine will calculate the
whole range using a Gauss-Legendre quadrature (we use 40 points).

\section{Thermally Averaged Compton Scattering Cross-Section}
To calculate the thermally averaged Compton scattering cross-section
$\sigma(x)\equiv \sigma_{CS}(x)/\sigma_{T}$, we use the full
integral expression in PS96.  However, that expression fails numerically
for low corona temperatures $\Theta\equiv kT_{e}/(m_ec^2)$
and low photon energies
$x$.  For low $\Theta$, the modified Bessel function $K_{2}(1/\Theta)$ used in
the normalization underflows, while for low $x$, the integrand cannot be
calculated accurately due to high order cancellation of the $2/x$ term.
We therefore switch to various asymptotic expressions for low $\Theta$
or low $x$ as needed.

For corona temperatures below $10^{7}$ K, we use
\begin{equation}
	\sigma(x)=1-2x+{\frac{26}{5}}x^{2} \qquad x\leq 10^{-3}
\end{equation}
and
\begin{equation}
	\sigma(x)={\frac{3}{8x^{2}}}\left[4+\left(x-2-\frac{2}{x}\right)
    \ln(1+2x)+\frac{2x^{2}(1+x)}{{(1+2x)}^{2}}\right] \qquad x>10^{-3},
\end{equation}
which are accurate to 1 part in $10^{7}$ in the ranges indicated.

For corona temperatures above $10^{7}$ K, we use
\begin{equation}
	\sigma(x)={\frac{1}{K_{2}(1/\Theta)}}\sum_{n=0}^{\infty}{(-2x)}^{n}
	a_{n}K_{n+2}(1/\Theta),
\end{equation}
where
\begin{equation}
a_{n}=\frac{3}{8}\left(n+2+
	\frac{2}{n+1}+\frac{8}{n+2}-\frac{16}{n+3}\right),
\end{equation}
for $x\leq x_{c}$,
where
\begin{equation}
	x_{c}=8.865351\times 10^{-4}{(1/\Theta)}^{0.353736486}.
\end{equation}
Otherwise, for $x>x_c$, we use the full integral expression of PS96.

Using 8 terms in the modified Bessel function series and 40-point
Gauss-Laguerre quadrature for the full integral expression, the error
in $\sigma(x)$ is less than 1 part in $10^{6}$
for corona temperatures above $10^{7}$ K.

\newpage

\figcaption[1]{Degree of polarization of radiation emerging from an
extremely low temperature corona with
$T_{e}=1.16$ K.  The input spectrum is a uniform,
unpolarized, low energy radiation field extending up to $E=0.332$ keV.
Open squares are from a $\tau =10.0$ corona
with a perfectly absorbing accretion disk surface and
isotropic sources at the corona base.
Other points are for a perfectly reflecting disk surface at the corona base
and a source having a $1/\mu$ intensity profile.
Crosses, filled squares, open triangles, and filled triangles are from coronae
of $\tau =10.0$, $2.0$, $1.0$, and $0.20$, respectively.
The solid line is the classical Chandrasekhar (1960)
result of polarization by Thomson scattering in an optically thick electron
slab.}

\figcaption[2]{Degree of polarization of radiation emerging from $\tau =10.0$
coronae with various temperatures.  The input spectrum is a uniform, isotropic,
unpolarized, low energy radiation field extending up to $E=0.332$ keV,
and the polarization profiles shown are calculated for the same
energy range.  Open squares, filled
squares, filled triangles, open triangles, crosses are from coronae
with $kT_{e}=1$, 10, 100, $10^3$, and $10^4$ keV, respectively.
The solid line is the classical Chandrasekhar
result of polarization by Thomson scattering in an optically thick electron
slab.}

\figcaption[3]{Cross-sections of the scattered intensity profile for
$kT_{e} = 1$~keV, purely linearly polarized incident radiation of $0.1$ keV,
and no change in the photon energy upon scattering.
The solid line is the surface of total intensity (polarized and
unpolarized), and the dotted line is the surface of polarized
intensity, which has the same polarization direction as the incident radiation.
Here, the dotted line lies on top of the solid line.}

\figcaption[4]{Same as Fig. 3, except $kT_{e}=100$~keV and the incident and
final photon energies are 10~keV.}

\figcaption[5]{Same as Fig. 3, except $kT_{e}=10$~MeV and the incident and
final photon energies are 1~MeV.}

\figcaption[6]{Degree of polarization of radiation emerging from
$kT_{e}=10$ keV coronae with various scattering optical depth.
The input spectrum is a uniform, isotropic,
unpolarized, low energy radiation field extending up to $E=0.332$ keV,
and the polarization profiles shown are calculated for the same
energy range.  Filled squares, filled triangles, open triangles, crosses
are from coronae
with $\tau=10.0$, $1.0$, $0.5$, and $0.2$, respectively.
Note that filled squares are the same as the filled squares in figure 2.}

\figcaption[7]{(a) Specific intensity and (b) degree of polarization $(Q_x/I_x)$
emerging from a $\tau=0.5$, $\Theta=0.20$ corona with an isotropic, unpolarized,
monoenergetic ($x=10^{-5}$) input spectrum.  The solid and dashed curves
correspond to $\mu=0.211$ and 0.789 viewing angles, respectively.  The heavy
curves represent the total emerging spectrum, while the numbered
pairs correspond to the lowest scattering orders.}

\figcaption[8]{Specific intensity and degree of polarization
emerging from a $\tau=0.5$, $kT_{e}=100$ keV
corona with an isotropic, unpolarized, quasi-monoenergetic
($E_{photon}\approx 5$ eV) input spectrum.  Open squares (dotted line),
filled squares (solid line), open triangles (short dashed line),
filled triangles (long dashed line)
correspond to $\mu=0.069$, $0.330$, $0.670$, and $0.931$ lines of sight,
respectively.}

\figcaption[9]{Specific intensity and degree of polarization emerging
along a line of sight with $\mu=0.069$ (i.e. edge-on view) from $\tau=0.5$
coronae with an isotropic, unpolarized, quasi-monoenergetic
($E_{photon}\approx 5$ eV) input spectrum.  Filled squares (solid line) and
open squares (dotted line)
correspond to coronae with electron temperatures of $kT_{e}=10$
and 100~keV, respectively.}

\figcaption[10]{Specific intensity and degree of polarization
emerging along $\mu=0.069$ from $kT_{e}=100$ keV
coronae with an isotropic, unpolarized, quasi-monoenergetic
($E_{photon}\approx 5$ eV) input spectrum.  Filled triangles (dashed line),
open squares (dotted line), and filled squares (solid line)
correspond to coronae with scattering optical depths of $\tau=1.0$, $0.5$,
and $0.3$, respectively.}

\figcaption[11]{Specific intensity and degree of polarization
emerging from a $\tau=0.5$, $kT_{e}=100$ keV
corona with an isotropic, unpolarized, continuum background plus
an emission line as the input spectrum at the base of the corona.
Open squares (dotted line),
filled squares (short dashed line), open triangles (long dashed line),
filled triangles (dot-dashed line)
correspond to $\mu=0.069$, $0.330$, $0.670$, and $0.931$ lines of sight,
respectively.  The solid line in the intensity plot is the
input spectrum at the base of the corona.  The polarization
of the input spectrum is zero everywhere.}

\figcaption[12]{Same as Fig. 11 but for an absorption line.}

\figcaption[13]{Specific intensity and degree of polarization
emerging at a viewing angle of $\mu=0.069$ from a $\tau=0.5$,
$kT_{e}=100$ keV
corona with an isotropic, unpolarized, continuum background plus
various emission lines as the input spectrum at the base of the corona.
Open squares (dotted line) are the same as the open squares in figure 10.
Open triangles (dashed line) and filled squares (solid line) have an original
emission line two and four times  as strong as the open squares, respectively.}

\figcaption[14]{Specific intensity and degree of polarization
emerging from a $\tau=0.5$, $kT_{e}=100$ keV
corona with an isotropic, unpolarized absorption edge spectrum
as the input spectrum at the base of the corona.
Open squares (dotted line),
filled squares (short dashed line), open triangles (long dashed line),
filled triangles (dot-dashed line)
correspond to $\mu=0.069$, $0.330$, $0.670$, and $0.931$ viewing angles,
respectively.  The solid line in the intensity plot is the
input spectrum at the base of the corona.  The polarization
of the input spectrum is zero everywhere.}

\figcaption[15]{Same as Fig. 14 but for an emission edge as the input spectrum.}

\figcaption[16]{Specific intensity and degree of polarization
emerging at a viewing angle of $\mu=0.069$ (edge-on view) from a $\tau=0.5$,
$kT_{e}=100$ keV
corona with an isotropic, unpolarized absorption edge spectrum
as the input spectrum at the base of the corona.
Open squares (dotted line) are the same as the open squares in figure 13,
for which the intensity ratio across the discontinuity is 1:1000.
Open triangles (short dashed line) and filled squares (long dashed line) have
intensity ratios of 1:100 and 1:10, respectively.
The solid lines in the intensity plot are the input spectra at the base of
the corona for all three cases.}

\figcaption[17]{Specific intensity and degree of polarization
entering at the base of the corona and emerging at the top of the corona
at a viewing angle of $\mu=0.211$ from a $\tau=0.5$, $kT_{e}=100$ keV
corona.
Open squares (dotted line) are the spectrum at the base of the corona, which
is produced by an accretion disk atmosphere calculation by Agol.
Filled squares (solid line) are the spectrum emerging at the top of the
corona after Compton scattering.}

\figcaption[18]{Same as Fig. 17 but for a viewing angle of $\mu=0.789$.}

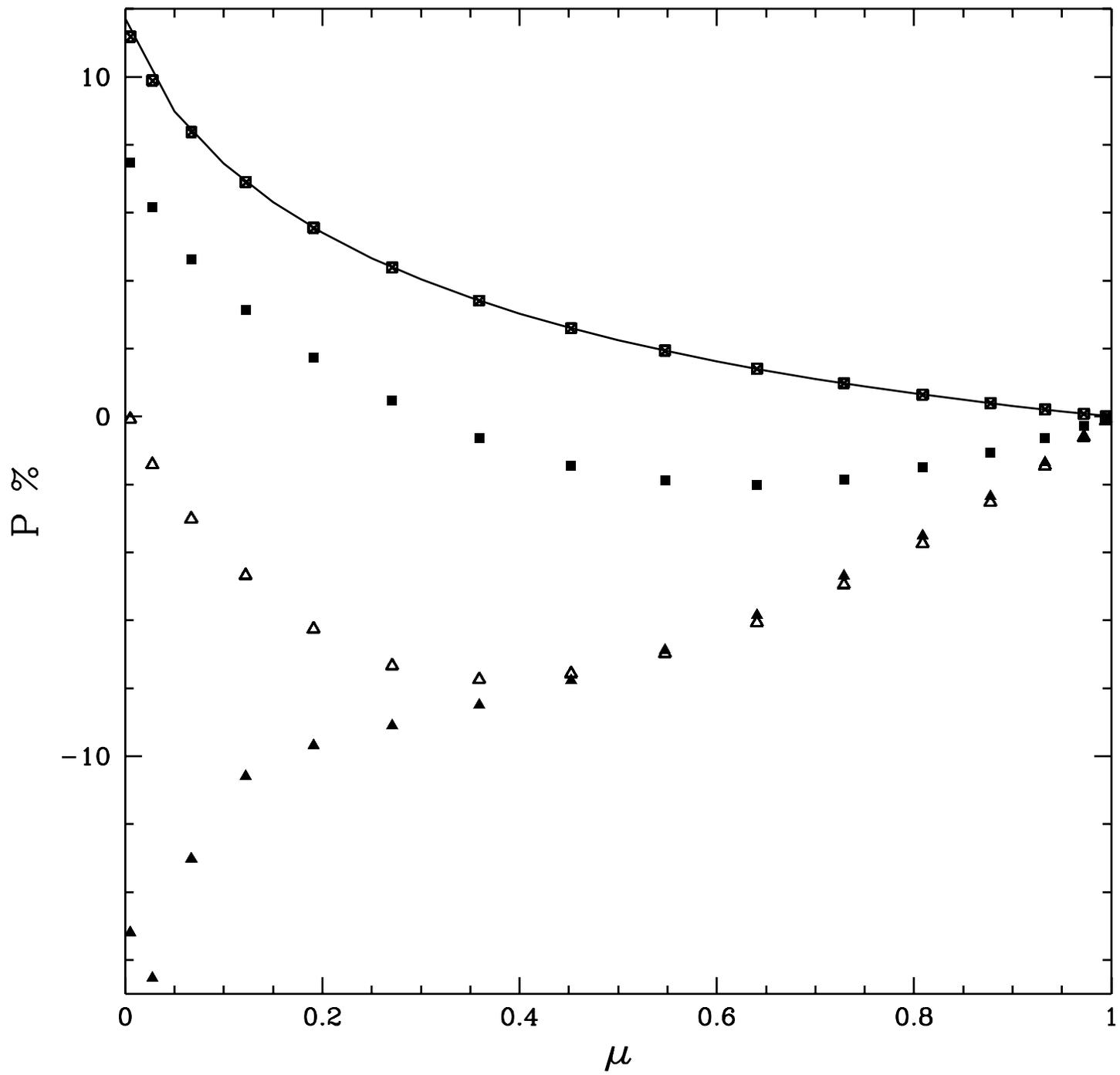
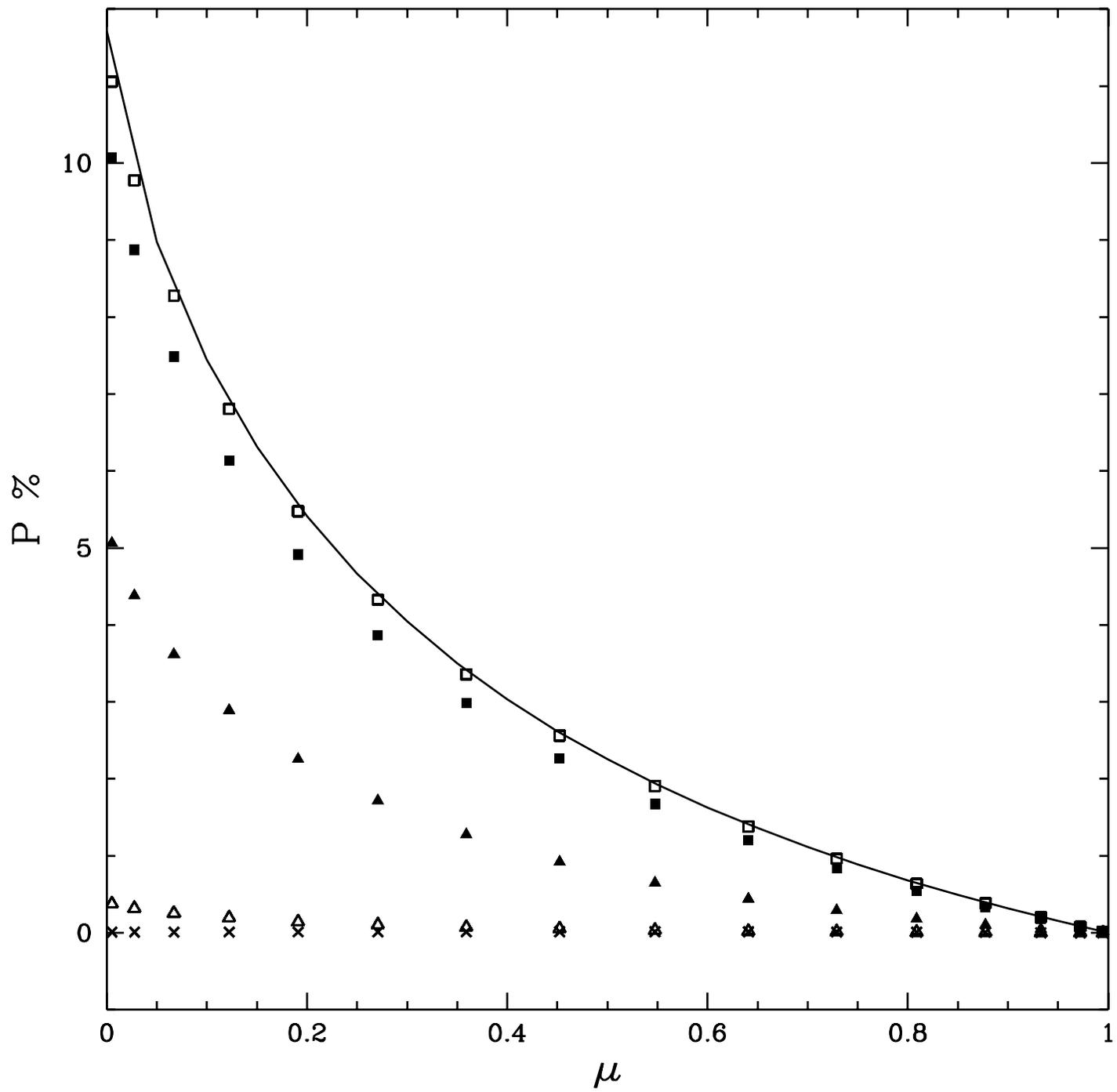
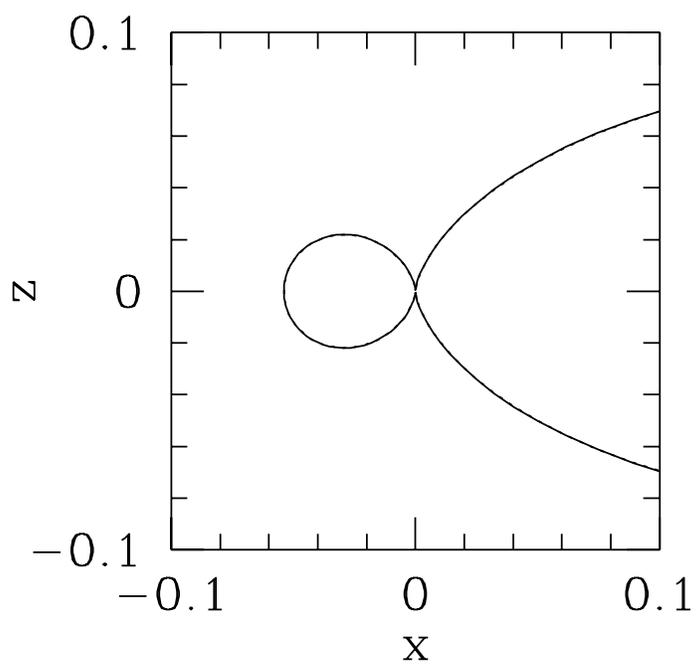
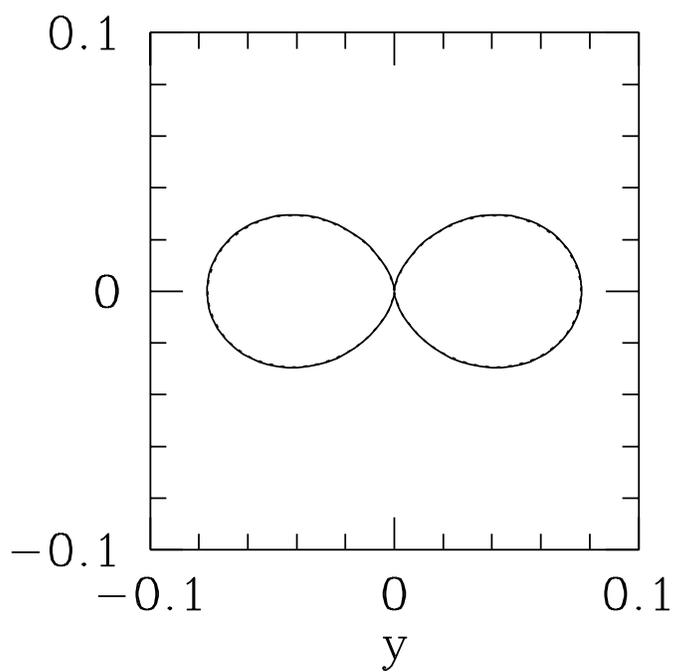
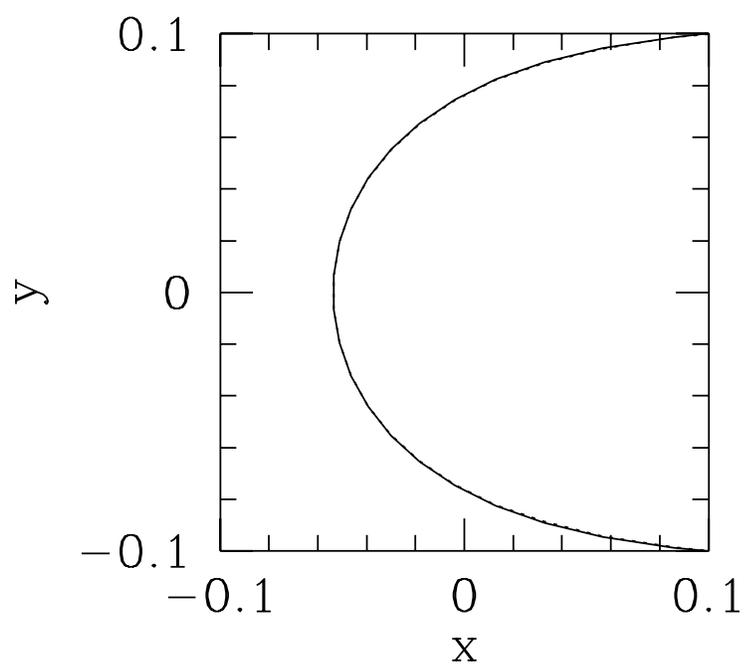
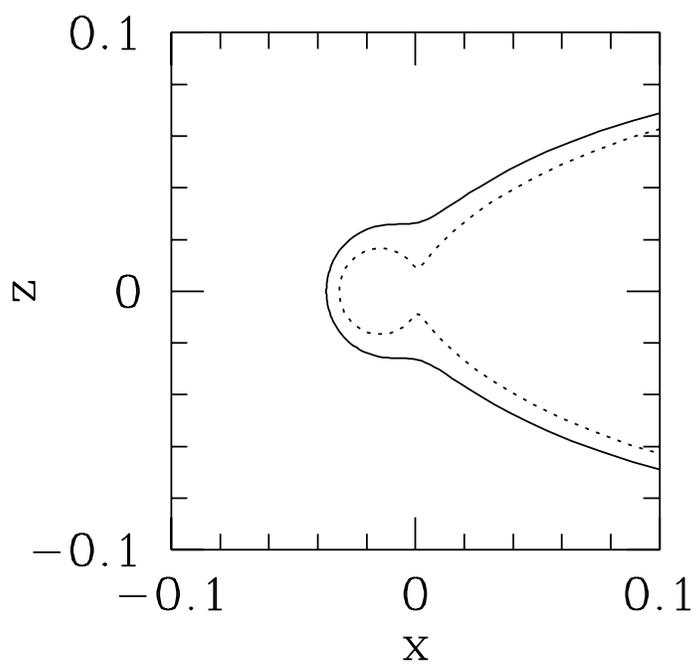
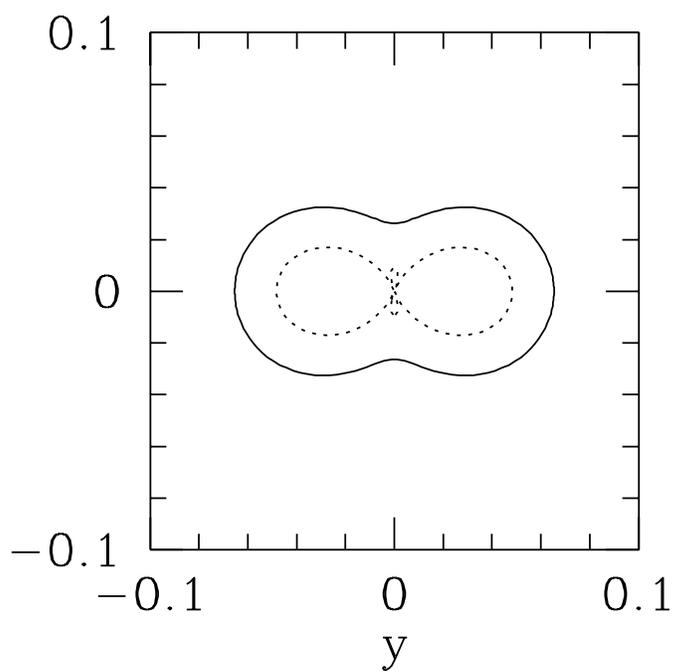
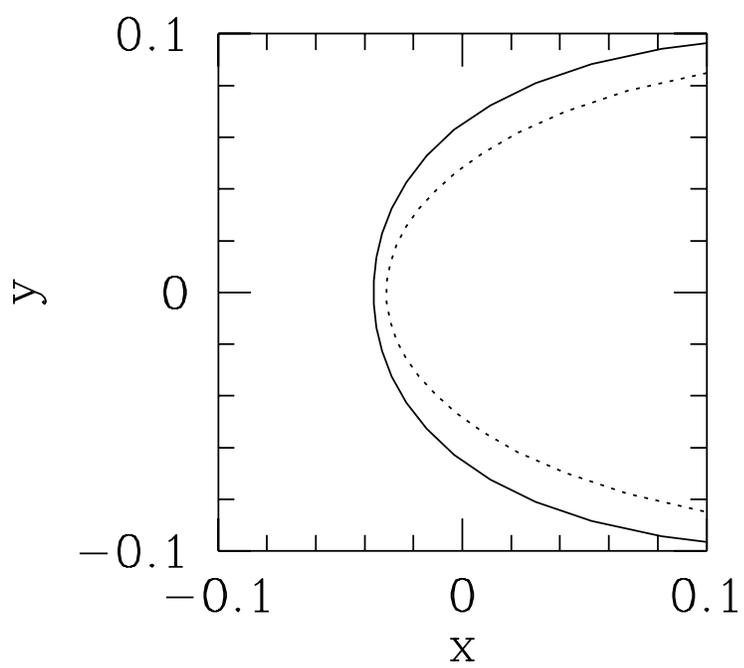
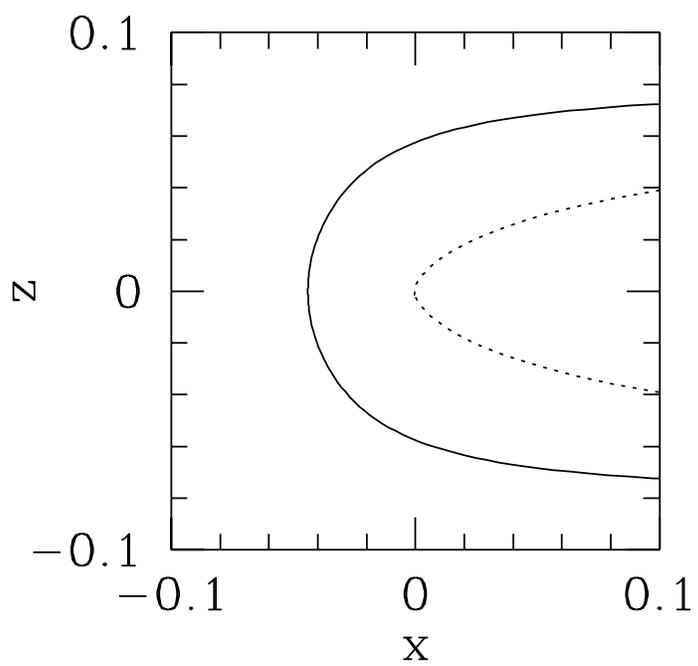
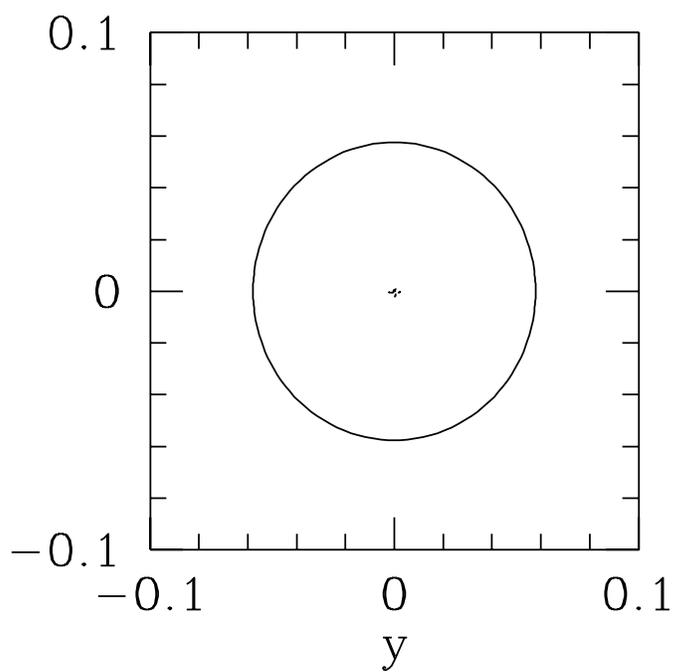
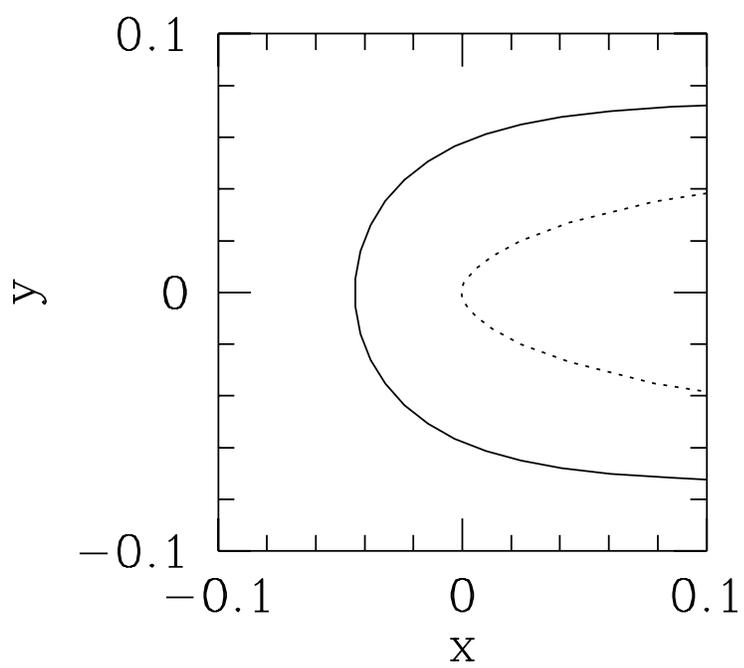
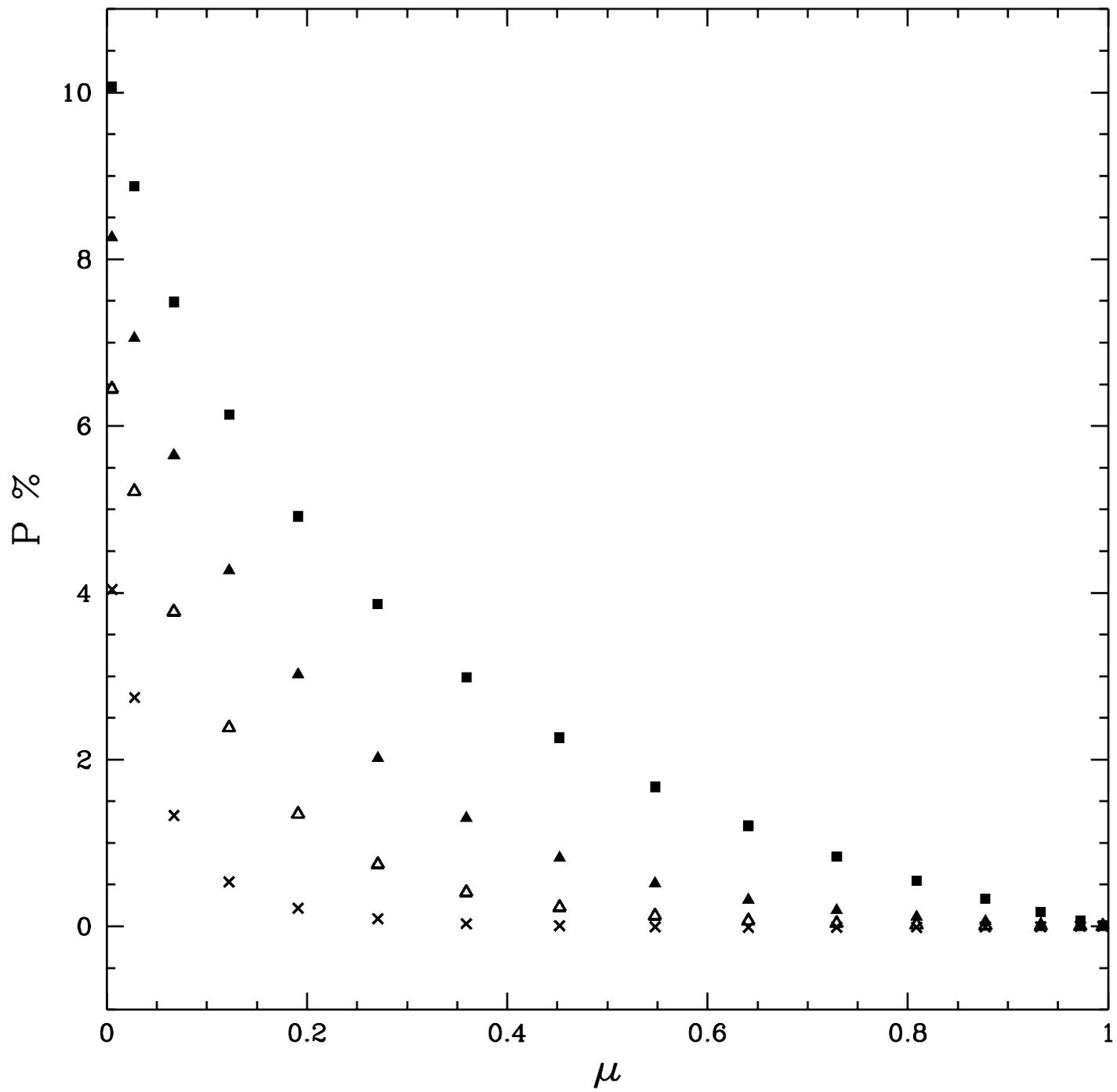
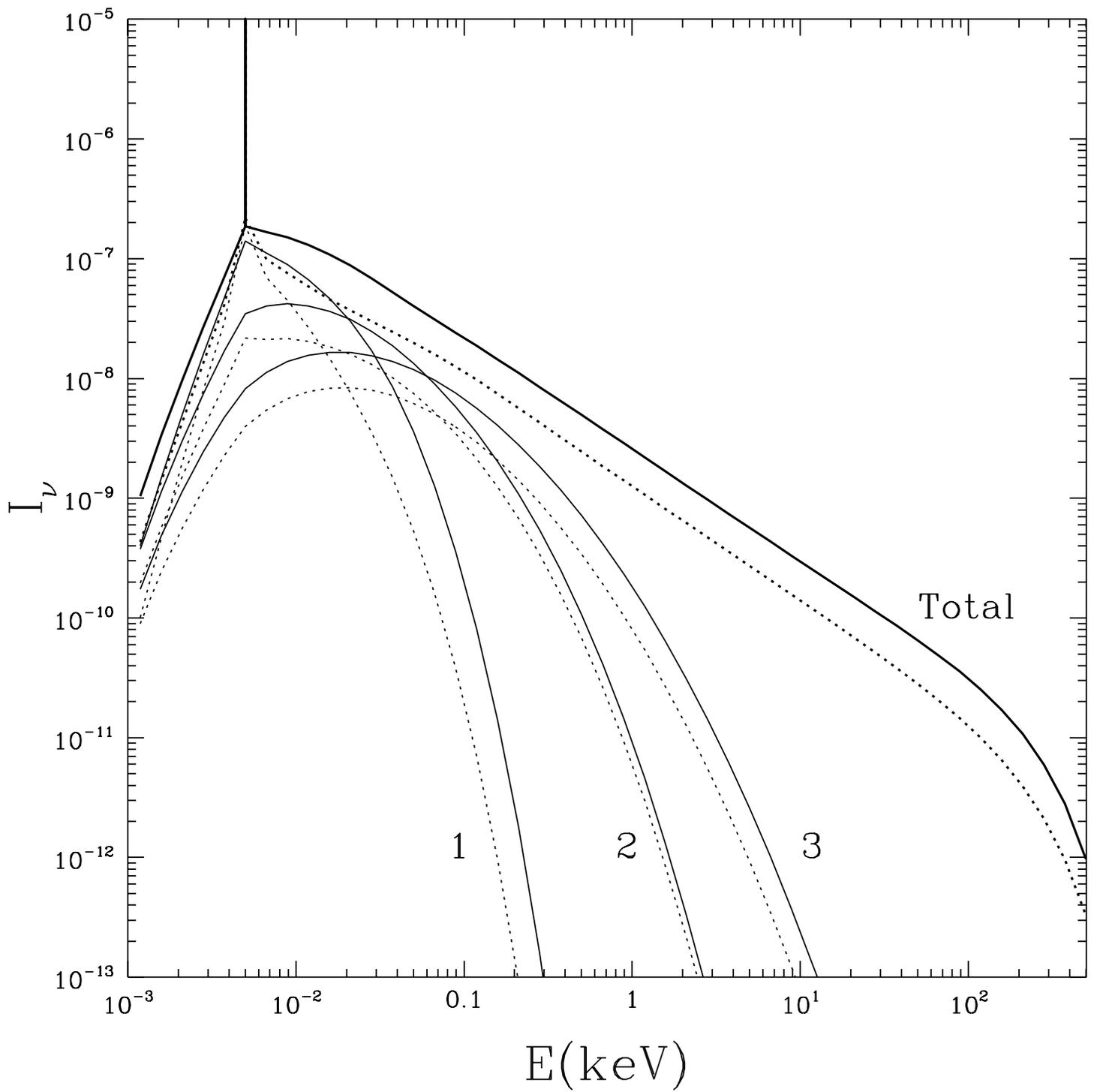
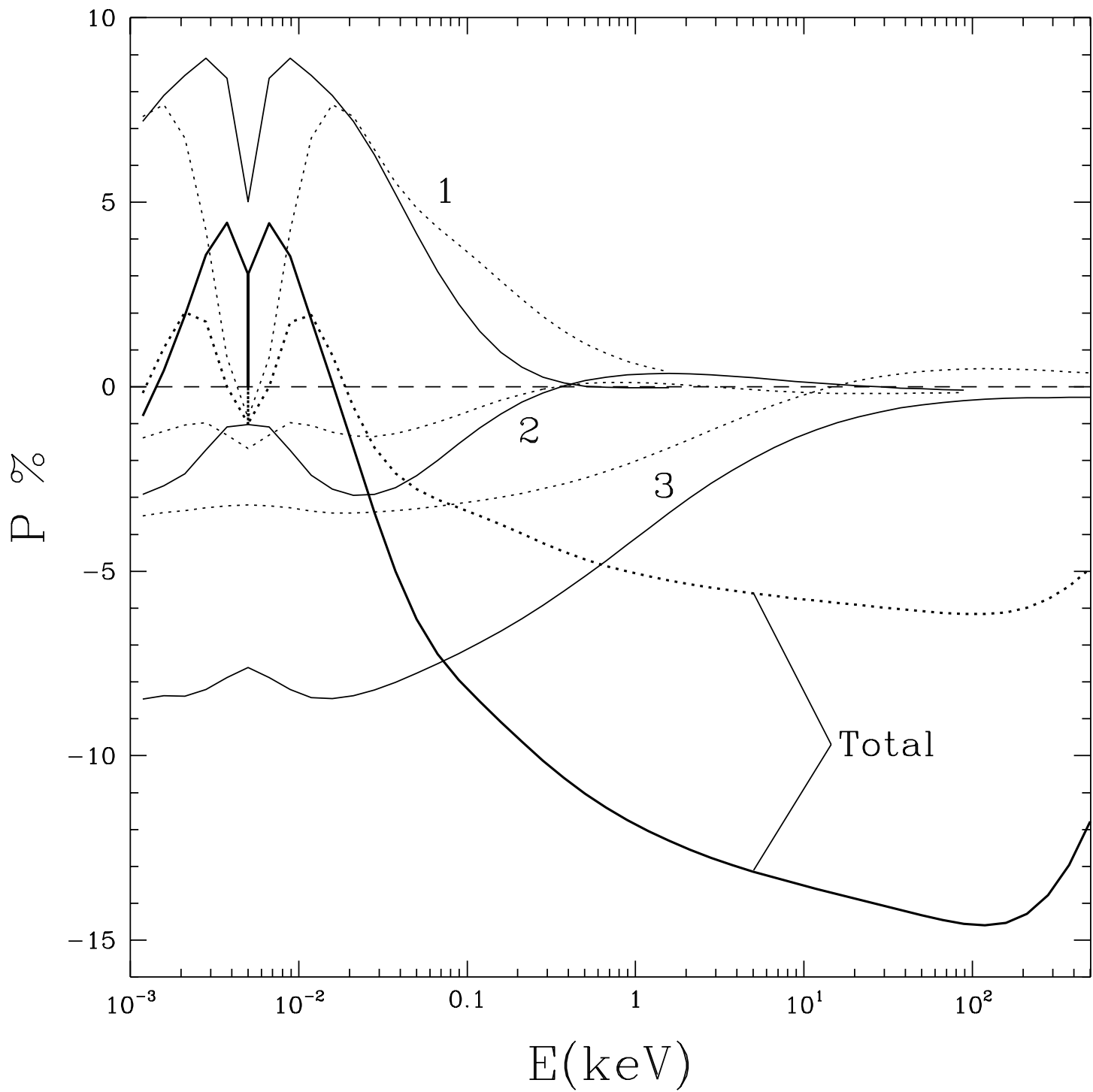
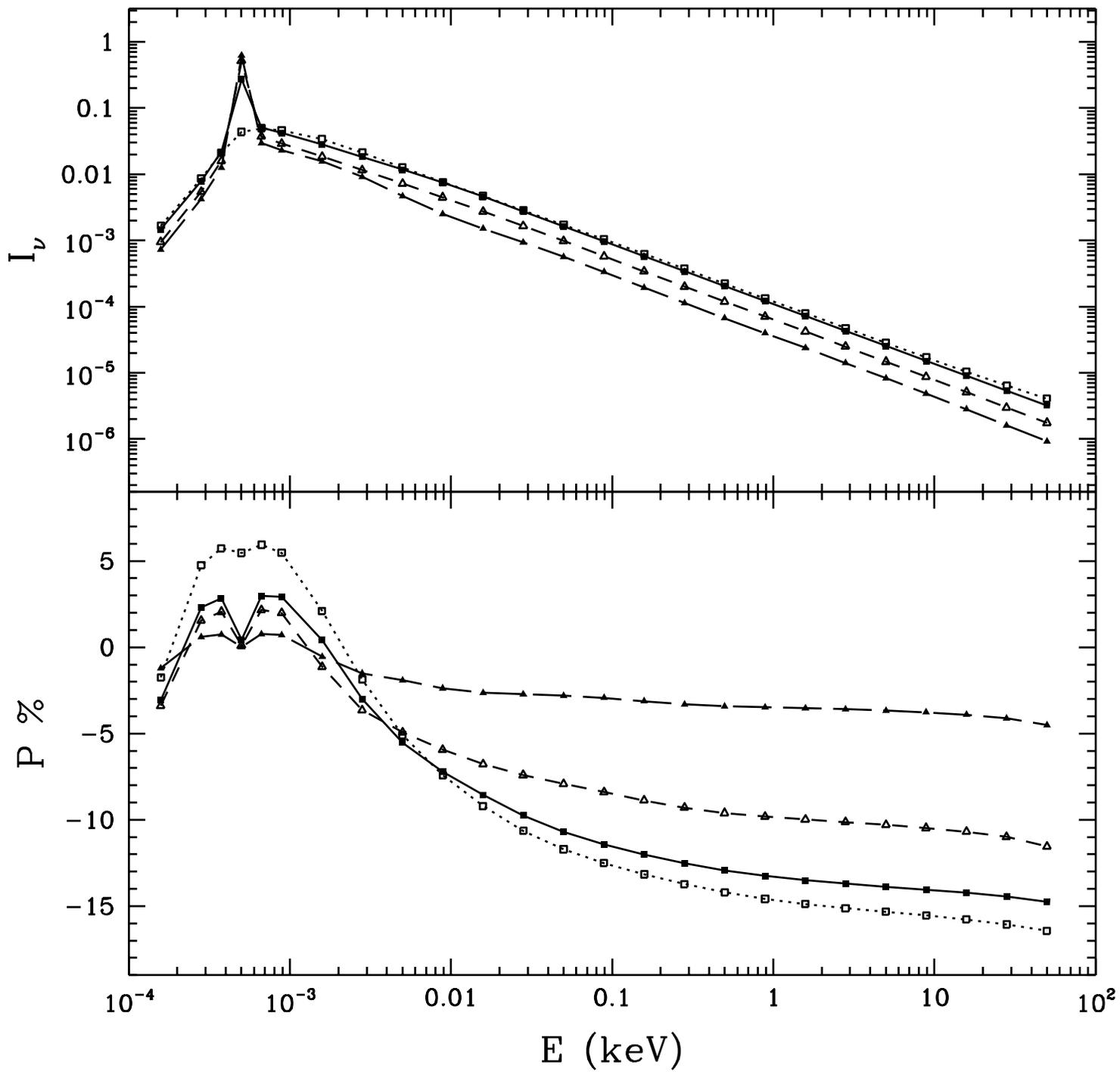
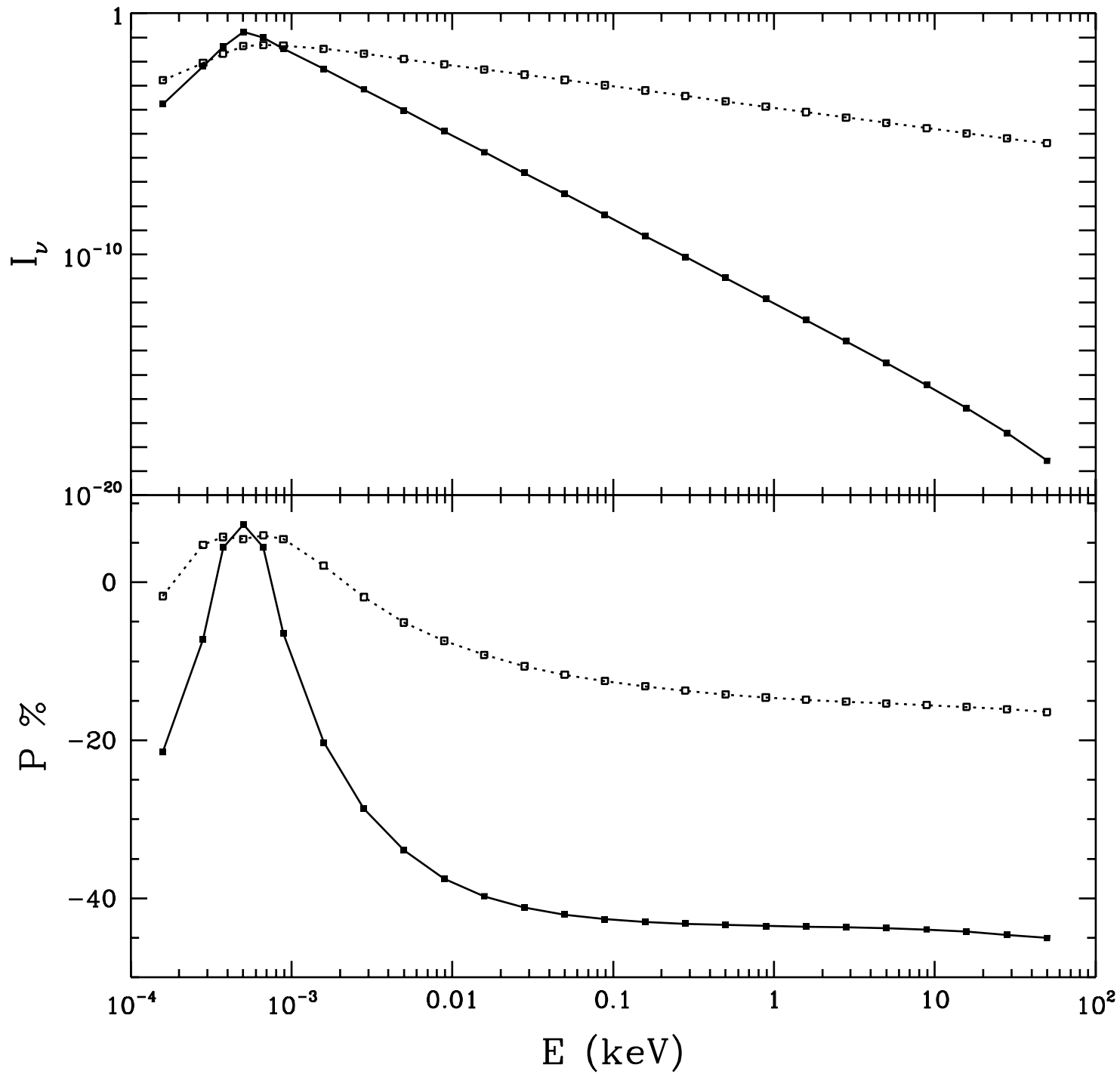
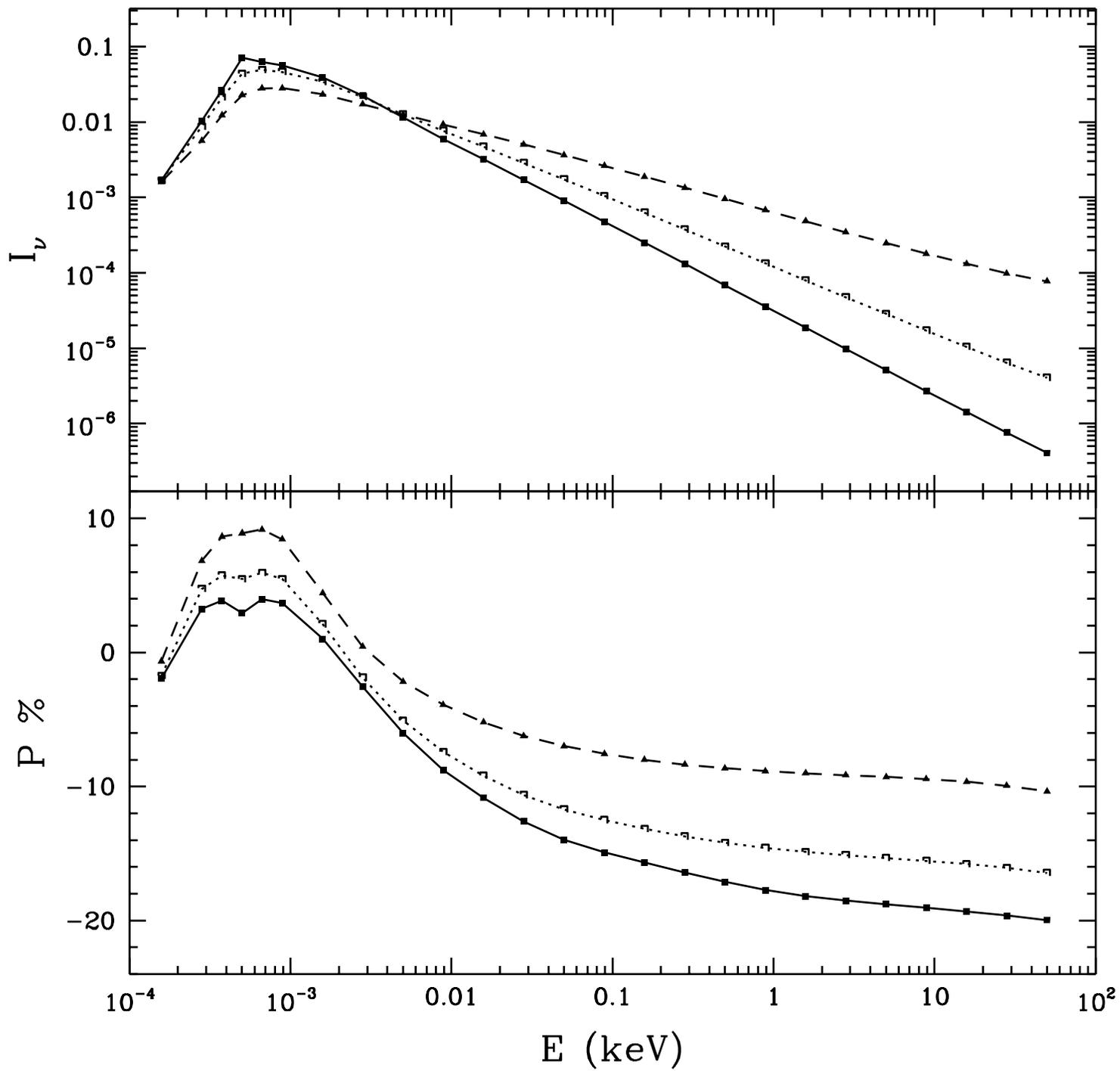
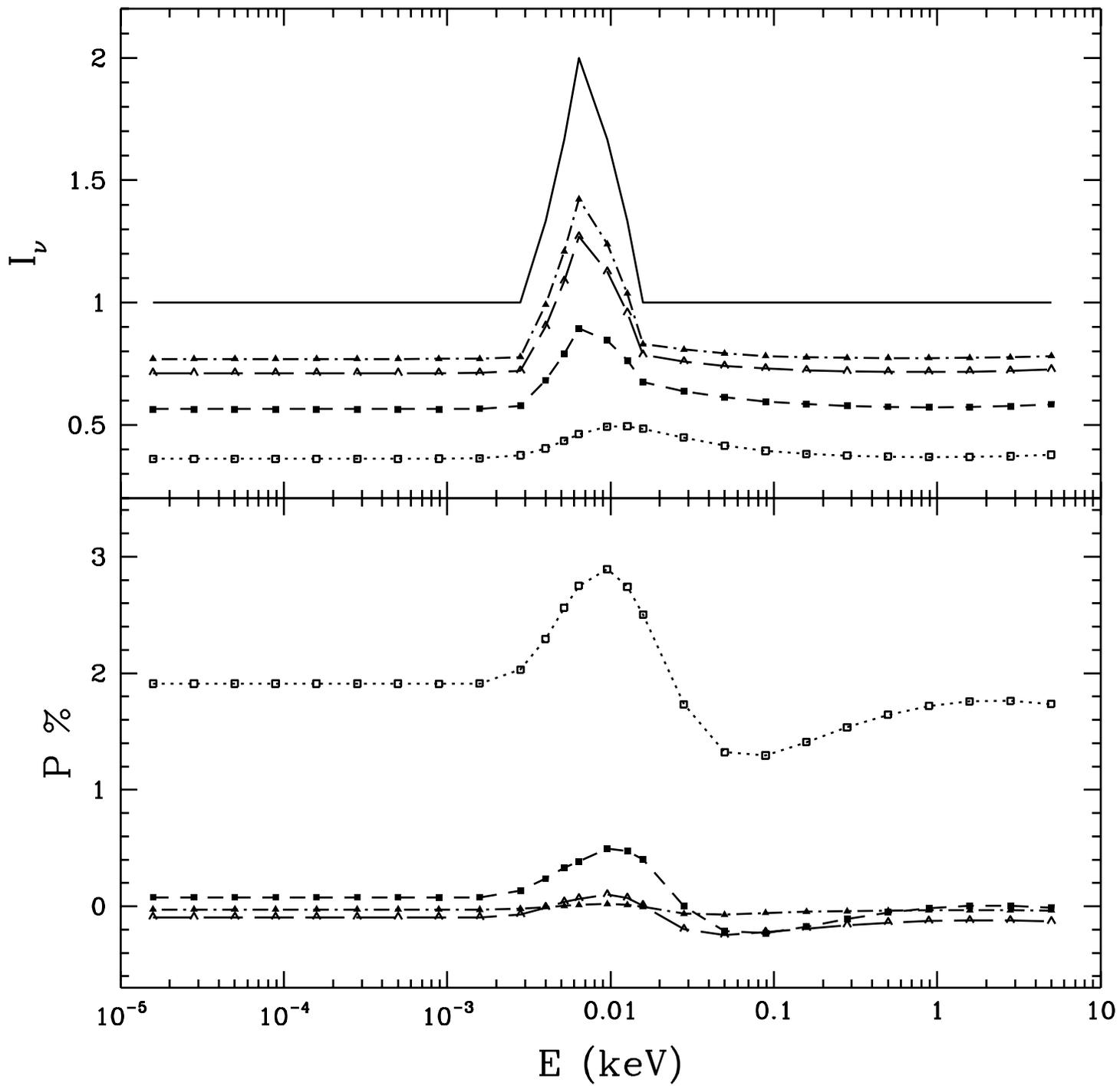
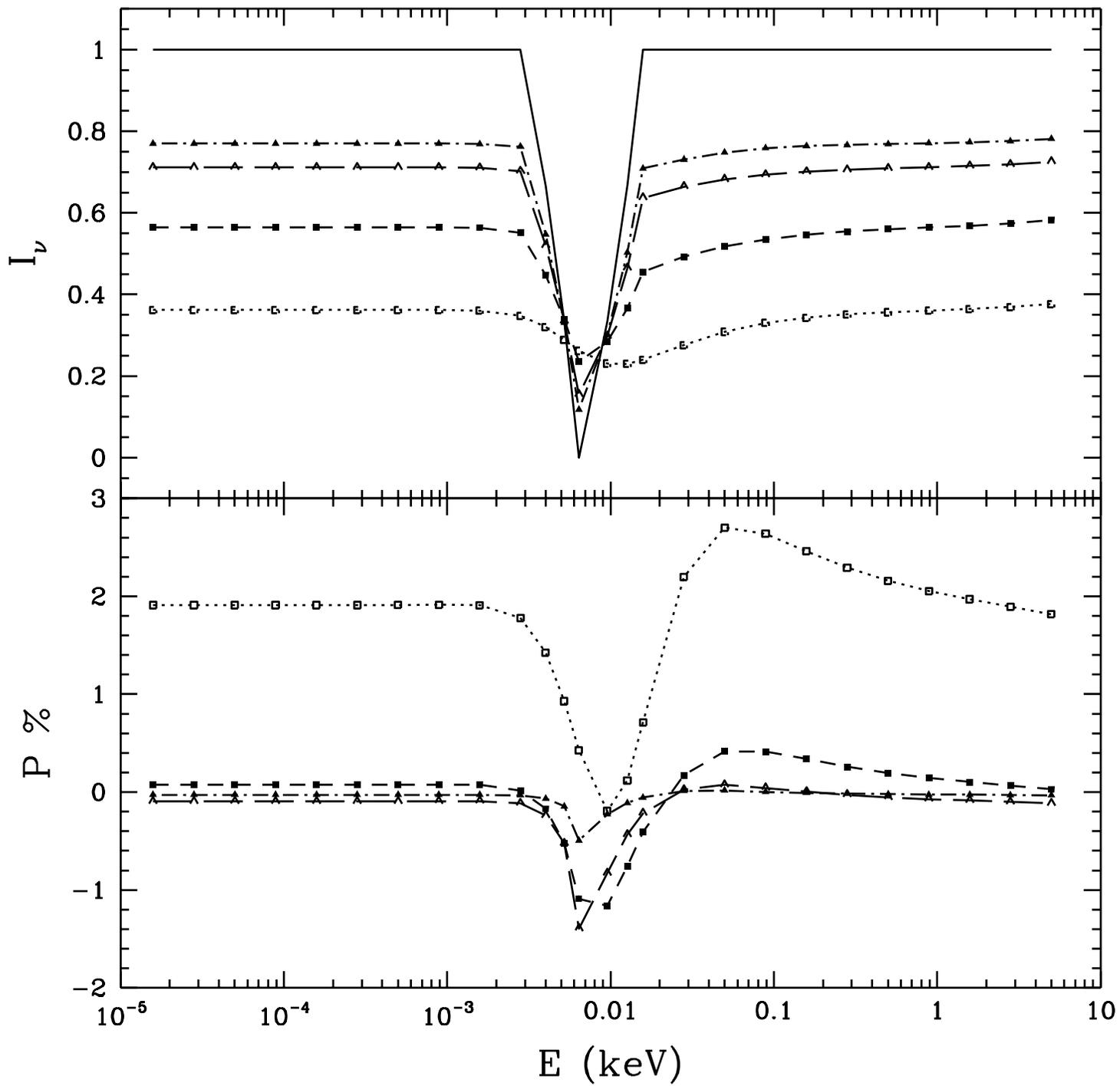
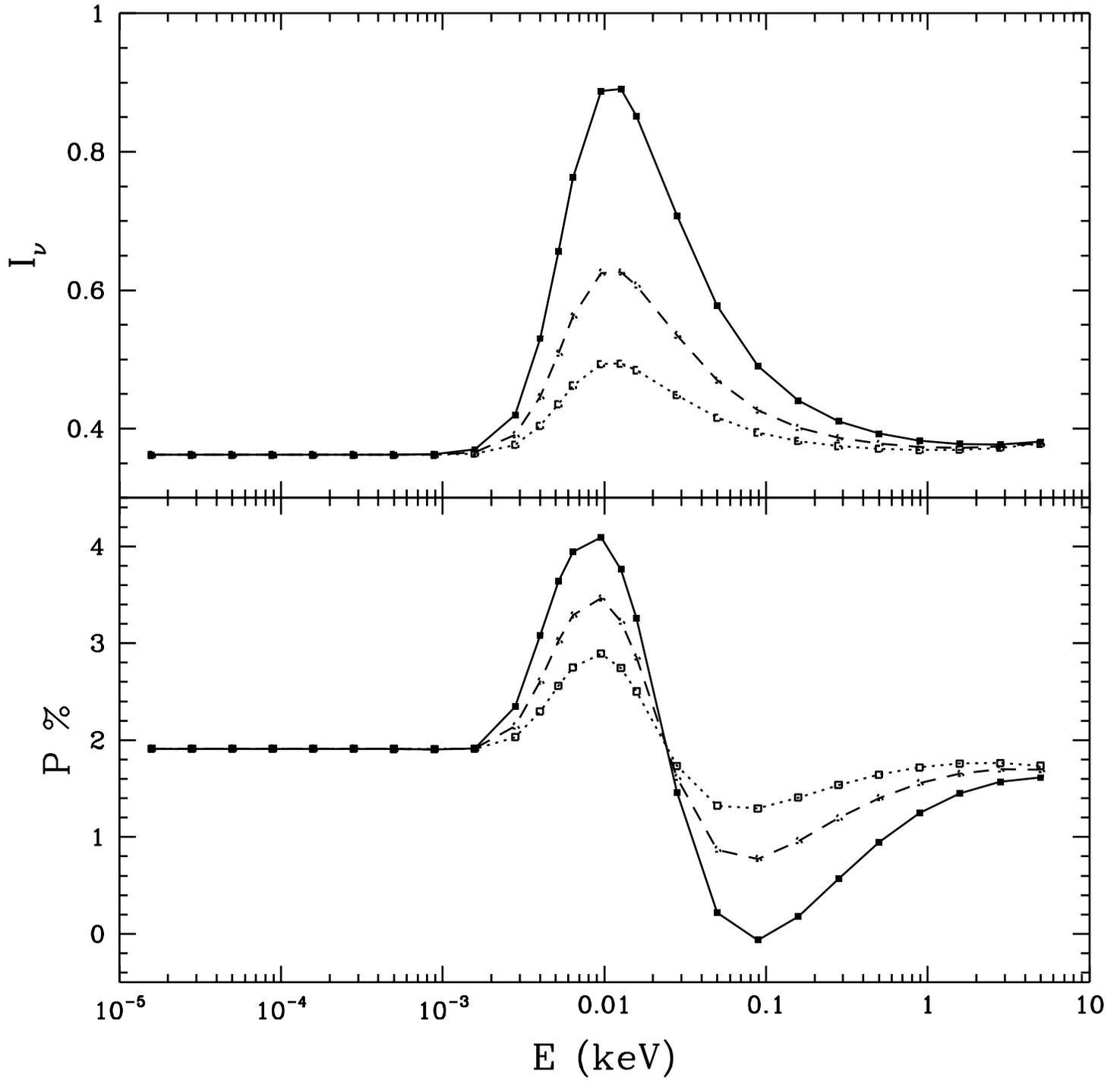
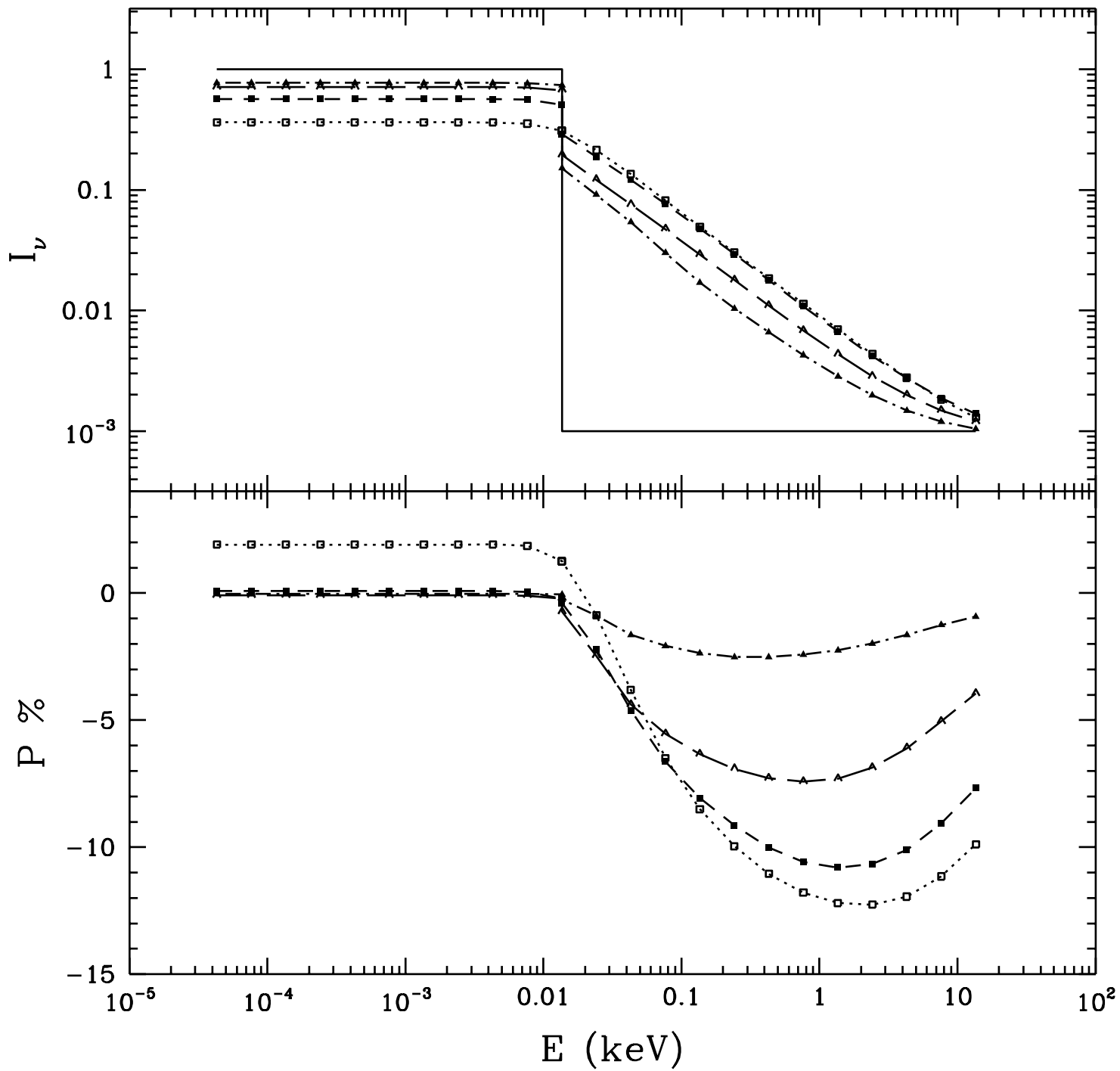
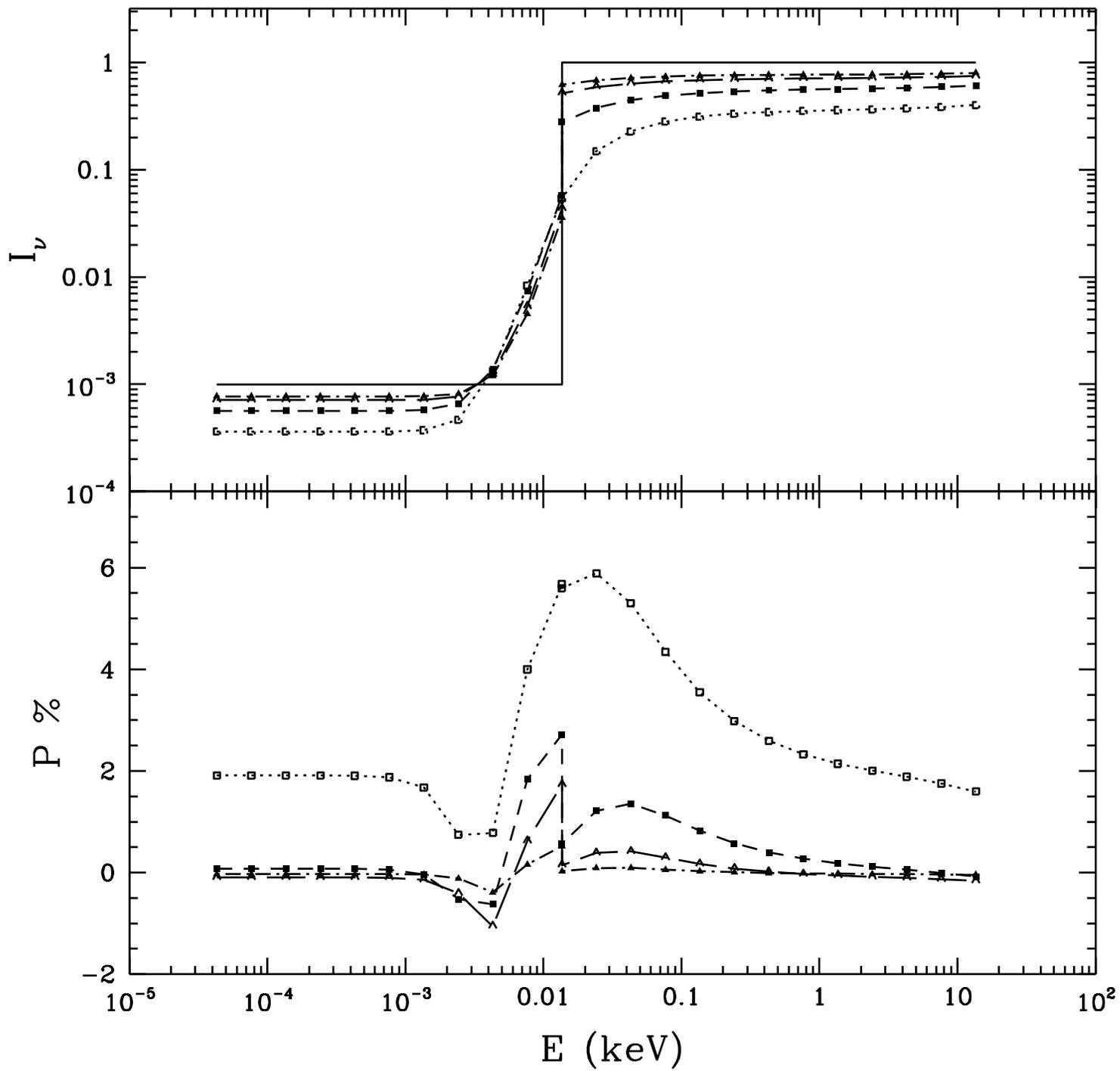
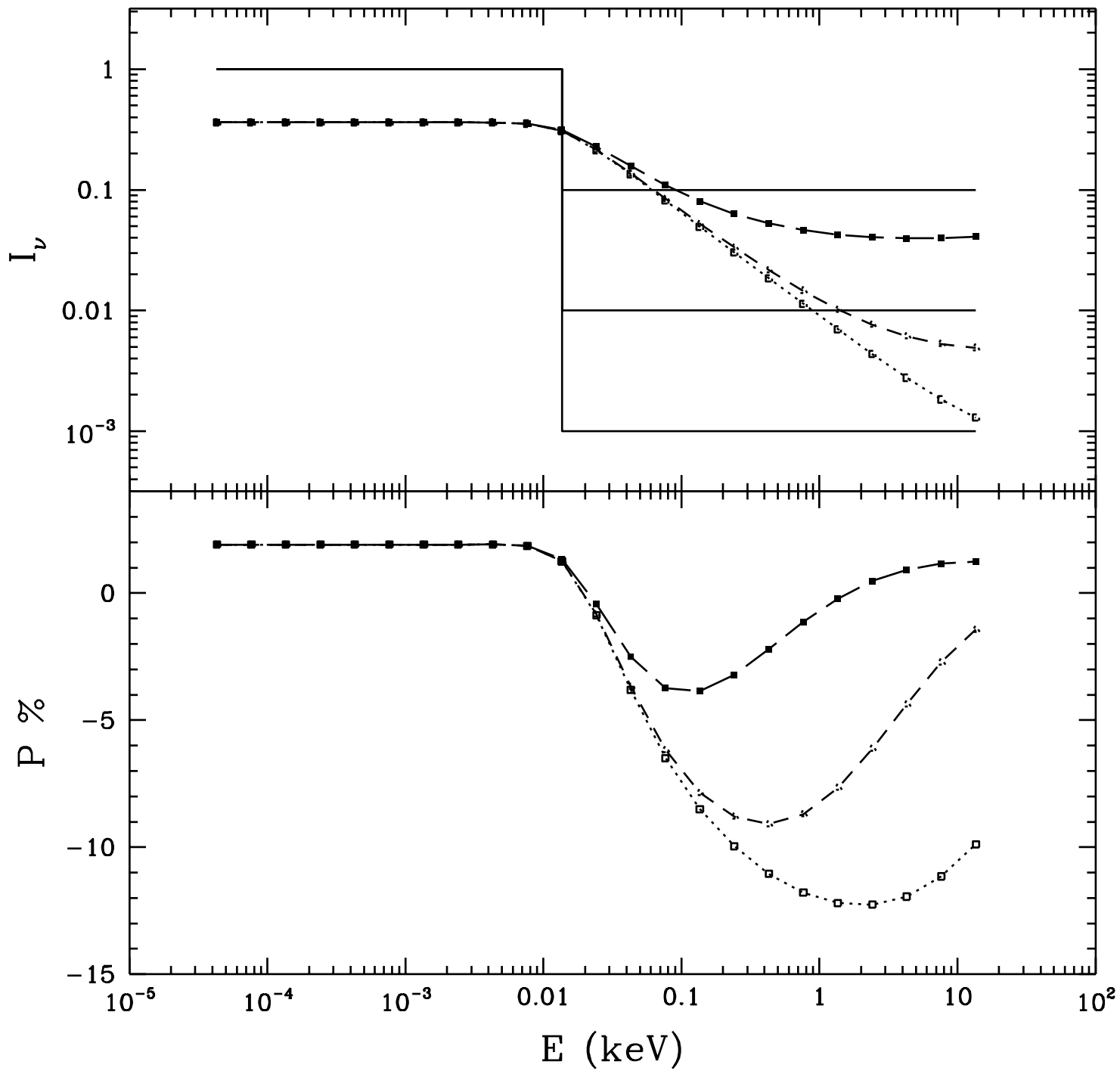
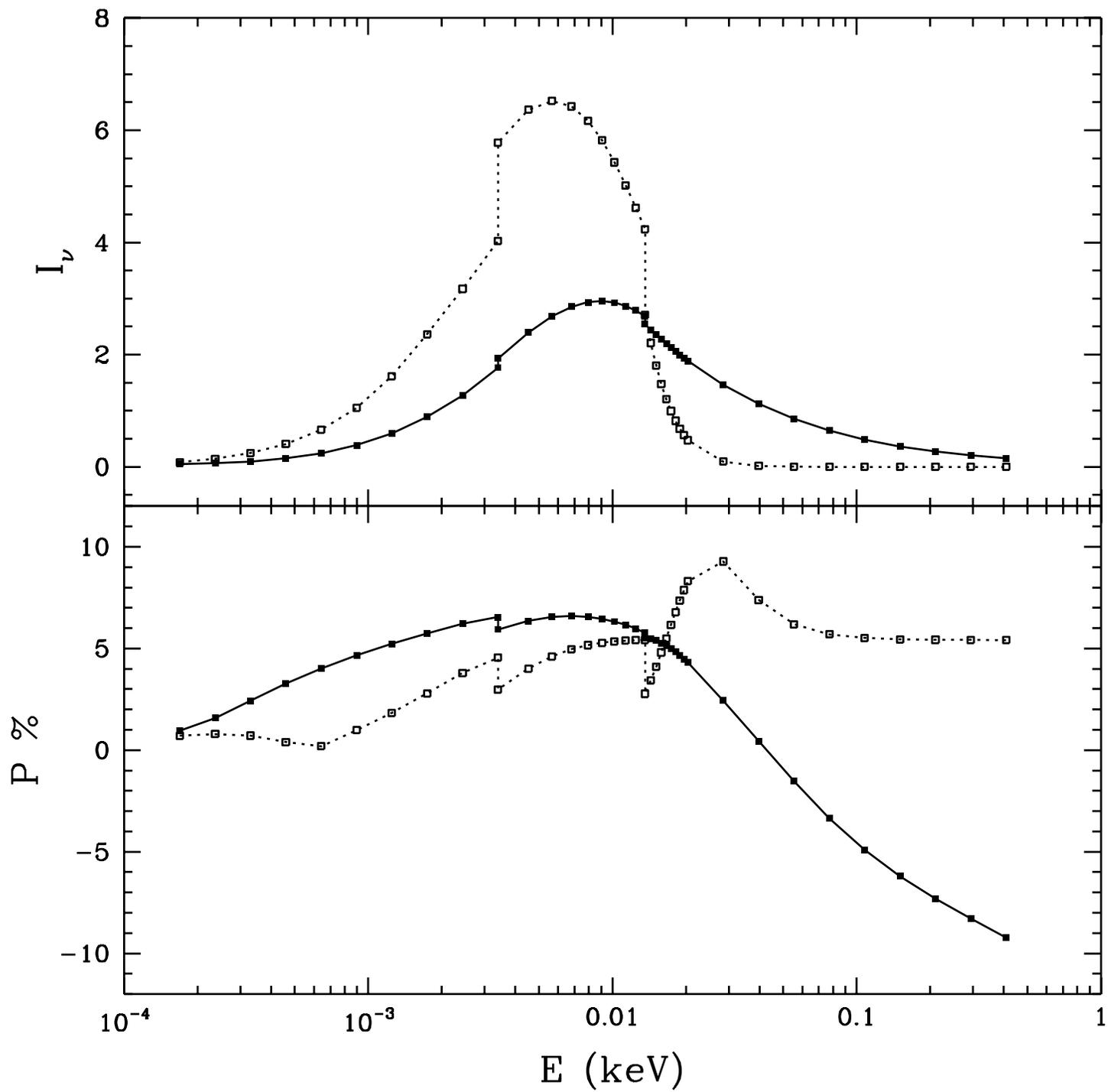
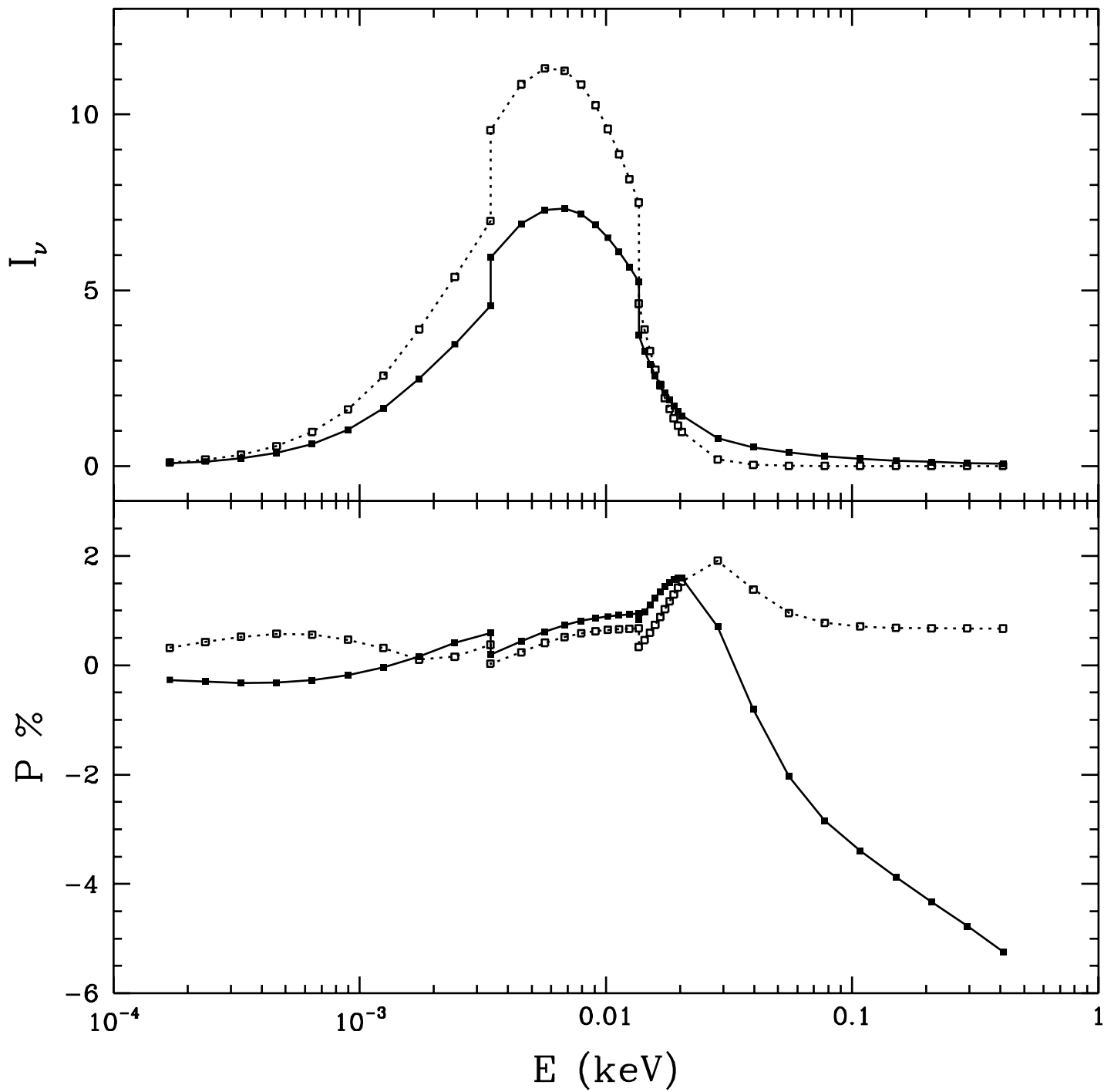
\end{document}